\documentclass[prd,nofootinbib,showpacs,preprintnumbers,amsmath,amssymb,superscriptaddress]{revtex4}
\usepackage{graphicx}
\usepackage{dcolumn}
\usepackage{bm}
\usepackage{epsfig}

\begin{document}

\preprint{USM-TH-249, arXiv:0907.0033 v2}

\title{Rational approximations in Analytic QCD}

\author{Gorazd Cveti\v{c}}
 
\email{gorazd.cvetic@usm.cl}

\affiliation{Department of Physics, Universidad T{\'e}cnica Federico
Santa Mar{\'\i}a, Valpara{\'\i}so, Chile}
\affiliation{Center of Subatomic Studies and Scientific-Technological Center of
Valpara{\'\i}so, Chile}

\author{H{\'e}ctor E. Mart{\'\i}nez}
\email{hector.martinez@usm.cl}

\affiliation{Department of Physics, Universidad T{\'e}cnica Federico
Santa Mar{\'\i}a, Valpara{\'\i}so, Chile}

\date{\today}

\begin{abstract}
\noindent
[In comparison with v1, in v2 Figs.6-8 are corrected due to a programming error;
analysis extended to two different IR cutoffs; Introduction rewritten; to appear in
J.Phys.G.] 

We consider the ``modified Minimal Analytic'' (mMA) coupling that
involves an infrared cut to the standard MA coupling. The mMA coupling 
is a Stieltjes function and, as a consequence, the paradiagonal Pad\'e
approximants converge to the coupling in the entire $Q^2$-plane
except on the time-like semiaxis below the cut. The
equivalence between the narrow width approximation of the
discontinuity function of the coupling, on the one hand, and this
Pad\'e (rational) approximation of the coupling, on the 
other hand, is shown. We approximate the analytic analogs of the
higher powers of mMA coupling by rational functions in
such a way that the singularity region is respected by the approximants.
Several comparisons, for real and complex arguments
$Q^2$, between the exact and approximate expressions are made and
the speed of convergence is discussed. Motivated by the success of
these approximants, an improvement of the mMA coupling is
suggested, and possible uses in the reproduction of experimental data 
are discussed.
\end{abstract}

\maketitle

\section{Introduction}
\label{sec:intr}


In perturbative QCD (pQCD) calculations, the coupling
$a(Q^2)\equiv \alpha_s(Q^2)/{\pi}$ shows non-physical
singularities at low energy ($-q^2\equiv Q^2 \lesssim 0.1 \ {\rm GeV}^2$).
The aim of analytic QCD (anQCD) is to give a coupling ${\mathcal{A}}_{1}(Q^2)$
which is analytic at low $Q^2$ $(>0)$ and reproduces the high
energy behavior of $a(Q^2)$. Using the Cauchy theorem we can write
QCD running coupling in the integral form
\begin{equation} 
a(Q^2)=\frac{1}{\pi}
\int_{\sigma=-\Lambda^2-\eta}^{\infty}d\sigma\frac{\rho_{1}(\sigma)}{\sigma+Q^2},
\label{pQCDdisp}
\end{equation} 
where $\eta\rightarrow +0$ and $\rho_{1}(\sigma)$ is the
discontinuity function of $a(Q^2)$ along the cut axis in the
complex $Q^2$-plane at $n$-loop approximation given by
$\rho_{1}^{(n-{\ell}.)}(\sigma)={\rm Im} [a^{(n-{\ell}.)}(-\sigma-i\epsilon)$].\\
The Minimal Analytic (MA) procedure of Shirkov and Solovtsov
\cite{Shirkov} removes the pQCD contribution of the unphysical
cut, $0 < -\sigma \leq \Lambda^2$, keeping the discontinuity
elsewhere unchanged
\begin{equation} \label{MAdisp}
{\mathcal{A}}_{1}^{(MA)}(Q^2)=
\frac{1}{\pi}\int_{\sigma=0}^{\infty}d\sigma\frac{\rho_{1}(\sigma)}{\sigma+Q^2}.
\end{equation}
This expression doesn't have singularities for $Q^2>0$ and, as
required, reproduces the high energy behavior of $a(Q^2)$. In
fact, ${\mathcal{A}}_{1}^{(MA)}(Q^2)$ is analytic in the entire complex plane
of $Q^2$ with the exception of negative semiaxis, reflecting the
analyticity properties of the space-like observables.

We will consider a modification to the MA coupling \cite{Shirkov}
of Eq.~(\ref{MAdisp}). Following Ref. \cite{Nesterenko}, the lower
limit of integration is increased to a certain value $M_{0}^{2}$
which we set to be $\sim M_{\pi}^{2}$ ($\sim 10^{-2}$-$10^{-1} \ {\rm GeV}^2$). 
Thus, the modified MA coupling is given by the following dispersion relation:
\begin{equation} 
{\mathcal{A}}_{1}^{(mMA)}(Q^2)=
\frac{1}{\pi}\int_{M_{0}^{2}}^{\infty}d\sigma\frac{\rho_{1}(\sigma)}{\sigma+Q^2}.
\label{mMAdisp}
\end{equation}
One of the motivations for modifying the MA in this way is to
include the point $Q^2=0$ and its vicinity
in the analyticity region of ${\mathcal{A}}_1(Q^2)$,
something reflected by analyticity properties of the space-like
observables, among them the V-type Adler function, where
the infrared (IR) cutoff $M_0^2$ is:
$M_0^2 \sim M_{\pi}^2$ ($\sim \Lambda^2_{\rm QCD}$).\footnote{
We note that $M_0$ is a renormalization scheme (RSch) dependent quantity,
just like $\Lambda$ is (e.g., $\Lambda$ in 
$V$-RSch and ${{\overline {\rm MS}}}$ RSch differ);
$s_0 \equiv M_0^2/\Lambda^2$ is RSch-independent. Therefore,
while we expect $M_0 \sim M_{\pi}$, $M_0$ ($\Leftrightarrow \ s_0$)
is a free parameter of the model.}
When $M_0^2 > 0$, the analytic coupling
${\mathcal{A}}_{1}$ is a Stieltjes function, with the
radius $R$ of convergence for its Taylor expansion around
$Q^2=0$ given by $R=M_0^2$. Stieltjes functions
have the attractive property that their paradiagonal
Pad\'e approximants must converge to such functions
at any complex value of the argument (except on the cut)
when the order index of the approximants increases. Such
a behavior is not guaranteed in MA model because there
$M_0^2=0$ and, consequently, $Q^2=0$ is not a point of
analyticity and ${\mathcal{A}}_{1}^{(MA)}$ is not a Stieltjes
function.

In this work we investigate the behavior of the
coupling ${\mathcal{A}}_{1}^{(mMA)}(Q^2)$ 
(with $M_0^2 \sim M_{\pi}^2$) of the mMA model (\ref{mMAdisp})
and its paradiagonal Pad\'e approximants. In Sec.~\ref{sec:PAandSF},
the basic definitions of rational (Pad\'e) approximants
and of Stieltjes functions are presented. In Sec.~\ref{sec:A1PA}
we demonstrate that the narrow width approximations for the
discontinuity function $\rho_1(\sigma)$ of
${\mathcal{A}}_{1}^{(mMA)}$ are equivalent to
approximating ${\mathcal{A}}_{1}^{(mMA)}(Q^2)$
by paradiagonal Pad\'e approximants; we investigate
the convergence of the latter approximants to the
``exact'' function values  ${\mathcal{A}}_{1}^{(mMA)}(Q^2)$
for $Q^2 > 0$. In Sec.~\ref{sec:AnPA} we extend our analysis
to the higher power analogs ${\mathcal{A}}_{n}^{(mMA)}$
of $a^n$  ($n \geq 2$), construct the approximants for
${\mathcal{A}}_{n}^{(mMA)}$ based on the aforementioned
paradiagonal Pad\'e approximants for ${\mathcal{A}}_{1}^{(mMA)}$
and investigate their behavior for $Q^2 > 0$.
In Sec.~\ref{sec:CAs} we investigate the behavior of 
the aforementioned approximants for complex values of $Q^2$.
In Sec.~\ref{sec:expd} we propose an extension (improvement) of the
mMA model, motivated by the success of the narrow width
approximations of the discontinuity function $\rho_1(\sigma)$
in mMA; further, we point out the prospects for applications
of such simple analytic QCD models to fitting experimental
data. Section \ref{sec:summ} represents the summary of 
our results.

\section{Pad{\'e} Approximants and Stieltjes functions}
\label{sec:PAandSF}

Consider a function $f(z)$, of a complex variable $z$, with power
expansion about the origin given by
\begin{equation} \label{fexp}
f(z)\approx\sum_{n=0}^{N+M}f_{n}z^{n}.
\end{equation}
The Pad{\'e} (or: rational) approximant, $R_{M}^{N}(z)$, 
is defined as the rational function
\begin{equation} \label{RMN}
R_{M}^{N}(z)=\frac{\sum_{n=0}^{N}a_{n}z^{n}}{\sum_{n=0}^{M}b_{n}z^{n}},
\end{equation} 
satisfying the condition that its expansion about $z=0$ matches
$N+M+1$ terms in the power series expansion in (\ref{fexp}). The
paradiagonal Pad{\'e} $R_{M}^{M-1}$ can be written (using partial
fractions) in the following way:
\begin{equation} \label{RMMm1}
R_{M}^{M-1}(z)=\sum_{n=1}^{M}\frac{d_n}{z+z_n},
\end{equation}
where $z_n$ are zeros of the denominator in Eq.~(\ref{RMN}).

If the function $f(z)$ is a Stieltjes function with a finite
radius of convergence $R>0$, i.e.
\begin{equation} \label{fdisp}
f(z)=\frac{1}{\pi} \int_{0}^{1/R}d\tau\frac{g(\tau)}{1+\tau z},
\end{equation} 
with $g(\tau)$ a nonnegative function, thus $f(z)$ 
having the real coefficients of its series expansion around $z=0$ given by
\begin{equation} \label{fnint}
f_n=\frac{(-1)^{n}}{\pi} \int_{0}^{1/R}\tau^{n}g(\tau)d\tau,
\end{equation} 
then a strong convergence theorem in the theory of Pad{\'e} approximants 
applies which says: in the limit 
$M\rightarrow\infty$, $R_{M}^{M+J}(z)$ ($J \geq -1$) is equal to $f(z)$, 
and the poles of $R_{M}^{M+J}(z)$ are simple poles
which lie on the interval of the negative real axis given by
$-\infty < z < -R$ (Ref.~\cite{Pade1}, Sections 5.2, 5.4).

In particular, when $f(z)$ is a Stieltjes function, the Pad\'e
$R_M^{M-1}(z)$ of $f(z)$ has the form (\ref{RMMm1}) with $d_n >0$ and 
$z_n>0$, the poles $-z_n$ having the following ordering:
$-\infty < - z_{M} < -z_{M-1} < \cdots < - z_1 < - R$.

\section{Approximation of the coupling by rational functions}
\label{sec:A1PA}

In order to establish a relationship between ${\mathcal{A}}_{1}^{(mMA)}(Q^2)$
and a Pad{\'e} approximant, we first note that the modified MA
coupling in Eq.~(\ref{mMAdisp}) is a Stieltjes function, as defined
in Eq.~(\ref{fdisp}) by identifying: $Q^2 = z$,
$\sigma = 1/\tau$ and $\rho_1(\sigma) = \tau \tilde{g}(\tau)$. 
It has the convergence radius $M_{0}^2$, and
the series expansion about $Q^2=0$ given by
\begin{equation} \label{A1expQ}
{\mathcal{A}}_{1}^{(mMA)}(Q^2)= \frac{1}{\pi}
\int_0^{1/M_0^2}d\tau\frac{\tilde{g}(\tau)}{(1+\tau
Q^2)} =\sum_{n=0}^{\infty}\tilde{L}_{n}(Q^2)^n,
\end{equation} 
where
\begin{equation} \label{tLn}
\tilde{L}_{n}=
\frac{(-1)^{n}}{\pi}\int_{0}^{1/M_{0}^{2}}\tau^{n}
\tilde{g}(\tau)d\tau,
\end{equation} 
and $\tilde{g}(\tau)=\sigma\rho_{1}(\sigma)$.
For simplicity, we want the expansion coefficients to be
dimensionless. To fulfill this, we introduce dimensionless variables
$s=\sigma/\Lambda^{2}=1/t$ and $u=Q^2/\Lambda^2$ in
Eq.~(\ref{mMAdisp}), where $\Lambda^2=\Lambda^2_{\rm QCD}$ (in
${\overline {\rm MS}}$ convention; 
$\Lambda^2_{\rm QCD} \sim 10^{-1} \ {\rm GeV}^2$). 
Consequently, Eqs.~(\ref{A1expQ}) and (\ref{tLn}) obtain the form
\begin{equation} \label{A1expu}
{\mathcal{A}}_{1}^{(mMA)}(u\Lambda^2)=\frac{1}{\pi}
\int_0^{1/s_0} dt \ \frac{g(t)}{1+tu}
=\sum_{n=0}^{\infty}L_{n}u^n,
\end{equation} 
with
\begin{equation} \label{Ln}
L_{n}= \frac{(-1)^{n}}{\pi}\int_{0}^{1/s_{0}}t^{n}g(t)dt,
\end{equation} 
where $u=Q^2/\Lambda^2=z/\Lambda^2$, 
$s_{0}=M_0^2/\Lambda^2$, $L_n=\tilde{L}_n(\Lambda^2)^n$ are
dimensionless, and $g(t)\equiv \rho_1(\Lambda^2/t)/t$.

An important consequence of the fact that ${\mathcal{A}}_1^{(mMA)}$ is a
Stieltjes function is that the series in Eq.~(\ref{A1expQ}) only
converges within the finite disc in the $Q^2$-plane ($Q^2<M_0^2$),
elsewhere the series is divergent. However, we will see that,
because of the aforementioned theorem \cite{Pade1}, 
only few coefficients of this
divergent series are needed in order to evaluate
${\mathcal{A}}_{1}^{(mMA)}(Q^2)$ in regions well beyond the convergence disc,
via approximate analytic continuation in the form of Pad{\'e}
approximants $R_M^{M-1}(z)$.

Following Ref. \cite{Peris}, we can approximate $\theta(s)\equiv
g(t\!=\!1/s)\equiv s\rho_1(\sigma\!=\!\Lambda^2s)$ by a singular series
expansion
\begin{equation} \label{gapp}
\frac{s \rho_1(s \Lambda^2)}{\pi} \equiv \frac{g(1/s)}{\pi} \equiv
\frac{\theta(s)}{\pi}\approx
\sum_{n=1}^{M} f_{n}^{2}s_{n} \delta(s-s_{n})\ ,
\end{equation} where $s = \sigma/\Lambda^2$,
$s_n=M_n^2/\Lambda^2$ and $f_n^2$ are all positive dimensionless
quantities. In Ref. \cite{Peris}, this type of approximation was
applied to the spectral function of the vector channel vacuum
polarization function $\Pi_{V}(Q^2)$. It was motivated there by
the highly singular behavior of the vector spectral function in
the limit $N_c \rightarrow \infty$. In our case we aren't working
in that limit, but we note that the spectral function
$\rho_{1}(\sigma)$ of ${\mathcal{A}}_{1}^{(mMA)}(Q^2)$ is positive; although
$\theta(s)$ has finite values, such a function can be well
approximated as a sum of positively weighted Dirac deltas for the
purpose of integration. We will see that this approximation is
equivalent to approximating the mMA coupling ${\mathcal{A}}_{1}^{(mMA)}(Q^2)$
as the Pad{\'e} approximant $R_M^{M-1}(u)$.

Using the singular approximation (\ref{gapp}) for $\theta(s)$ in
Eq.~(\ref{A1expu}), we obtain
\begin{equation} \label{A1app}
{\mathcal{A}}_{1}^{(mMA)}(u\Lambda^2)\approx
\sum_{n=1}^{M}\frac{f_{n}^{2}}{u+s_{n}}=
\sum_{n=1}^{M}\frac{F_{n}^{2}}{Q^2+M_{n}^{2}}
\end{equation} 
where $F_n^2=f_n^2 \Lambda^2$. This relation is the same 
as $R_{M}^{M-1}(z)$ Pad{\'e} of ${\mathcal{A}}_{1}^{(mMA)}$
given in Eq.~(\ref{RMMm1}), with $z=u$, $z_n=s_n$ and $d_n=f_n^2$
\begin{equation} \label{A1R}
{\mathcal{A}}_{1}^{(mMA)}(u\Lambda^2)\approx
R_{M}^{M-1}(u) = \sum_{n=1}^{M}\frac{f_{n}^{2}}{u+s_{n}}=
\sum_{n=1}^{M}\frac{F_{n}^{2}}{Q^2+M_{n}^{2}}
\end{equation} 
Therefore, by approximating $\theta(s)$ by positive
Dirac deltas we obtain ${\mathcal{A}}_1^{(mMA)}$ written in a
(paradiagonal) Pad{\'e} form.
Further, the following relation holds \cite{Pade1}:
\begin{equation} \label{difA1R}
| {\mathcal{A}}_{1}^{(mMA)}(u\Lambda^2)-R_{M}^{M-1}(u) |
\ \leq \
K \biggl|\frac{\sqrt{s_0+u}-\sqrt{s_0}}{\sqrt{s_0+u}+\sqrt{s_0}}
\biggr|^{2M},
\end{equation} 
where $K$ is a constant. Therefore 
$R_M^{M-1}(u)\rightarrow {\mathcal{A}}_{1}^{(mMA)}(u\Lambda^2)$ 
when $M\rightarrow\infty$.

In our evaluations we will use the perturbative coupling given by
the solution of the two-loop renormalization group (RG) equation
\begin{equation} \label{a2l}
a^{(2-{\ell}.)}(u\Lambda^2)=\frac{-1}{c_1(1+W_{\mp 1}(z_{\pm}))},
\end{equation} 
where $W_{\mp 1}(z)$ is the Lambert function (branches $n=\mp 1$), 
and the argument $z_{\pm}$ is
given by
\begin{equation} \label{zpm}
z_{\pm}=\left(\frac{|u|^{-\beta_0/c_1}}{c_1e}\right)e^{i(\pm\pi-\phi\beta_0/c_1)},
\end{equation} 
where $Q^2=|Q^2|e^{i\phi}$, $|u|=|Q^2|/\Lambda^2$,
$\beta_0=(11-2 n_f/3)/4$, $c_1=\beta_1/\beta_0$,
$\beta_1=(102-38n_f/3)/16$. For $0< \phi < \pi$, the branch
$W_{-1}(z_{+})$ is chosen, for $-\pi<\phi<0$, the branch
$W_{+1}(z_{-})$ is chosen.
At low energies ($|Q^2|\lesssim 10$ $GeV^2$), the number of active
quark flavors is $n_f=3$.  For details see Ref. \cite{Gardi:1998qr}.
With the coupling given in Eq.~(\ref{a2l}), the discontinuity
function becomes
\begin{equation} \label{rho12l}
\rho_1^{(2-{\ell}.)}(s \Lambda^2)=
{\rm Im} \left[\frac{-1}{c_1(1+W_{+1}(z_{-}(s)))}\right],
\end{equation} 
with $\phi=-\pi$ and $|u|=\sigma/\Lambda^2 \equiv s$
in Eq.~(\ref{zpm}). These expressions are in fact 
valid at any $n$-loop level ($n \geq 2 $) in the 
't Hooft renormalization scheme.
\begin{figure}[htb]
\centering\epsfig{file=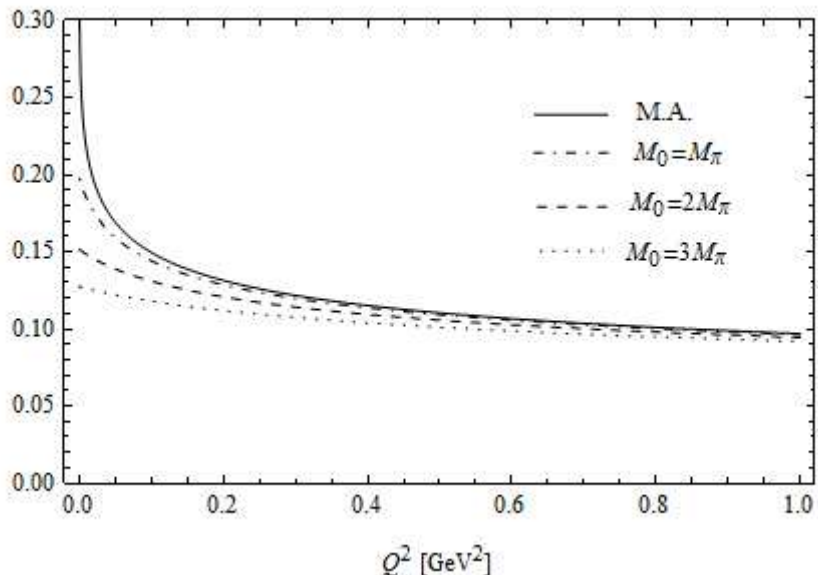,width=12.cm}
\vspace{-0.4cm}
\caption{${\mathcal{A}}_{1}^{(MA)}(Q^2)$ (continuous), and
${\mathcal{A}}_{1}^{(mMA)}(Q^2)$ at two-loop level for various
values of $M_0$ cut. Throughout this paper we 
use $n_f=3$ and $\Lambda_{{\overline {\rm MS}}} = 0.35$ GeV.}
\label{figure01}
\end{figure}
\begin{figure}[htb]
\begin{minipage}[b]{.49\linewidth}
\centering\epsfig{file=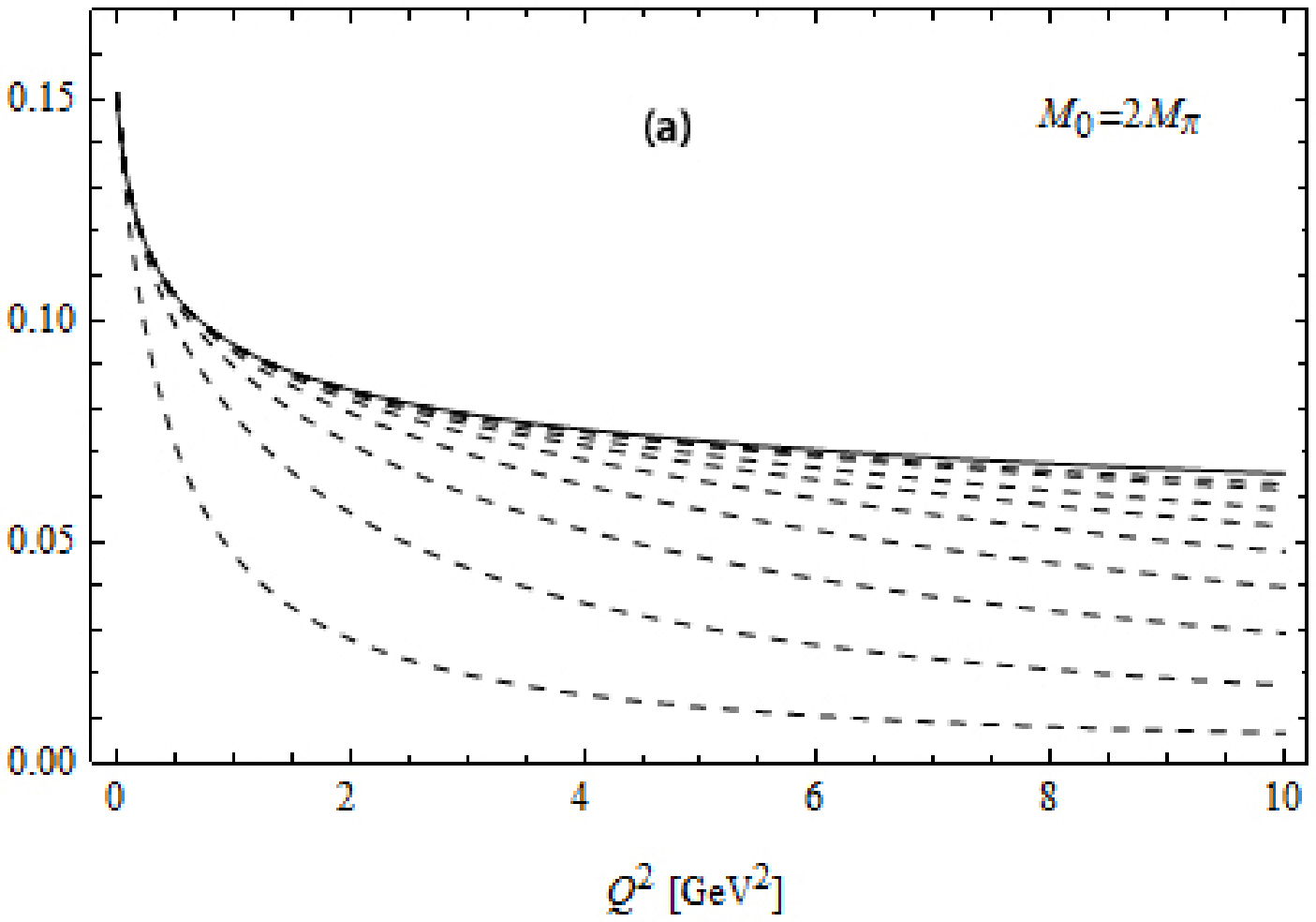,width=\linewidth}
\end{minipage}
\begin{minipage}[b]{.49\linewidth}
\centering\epsfig{file=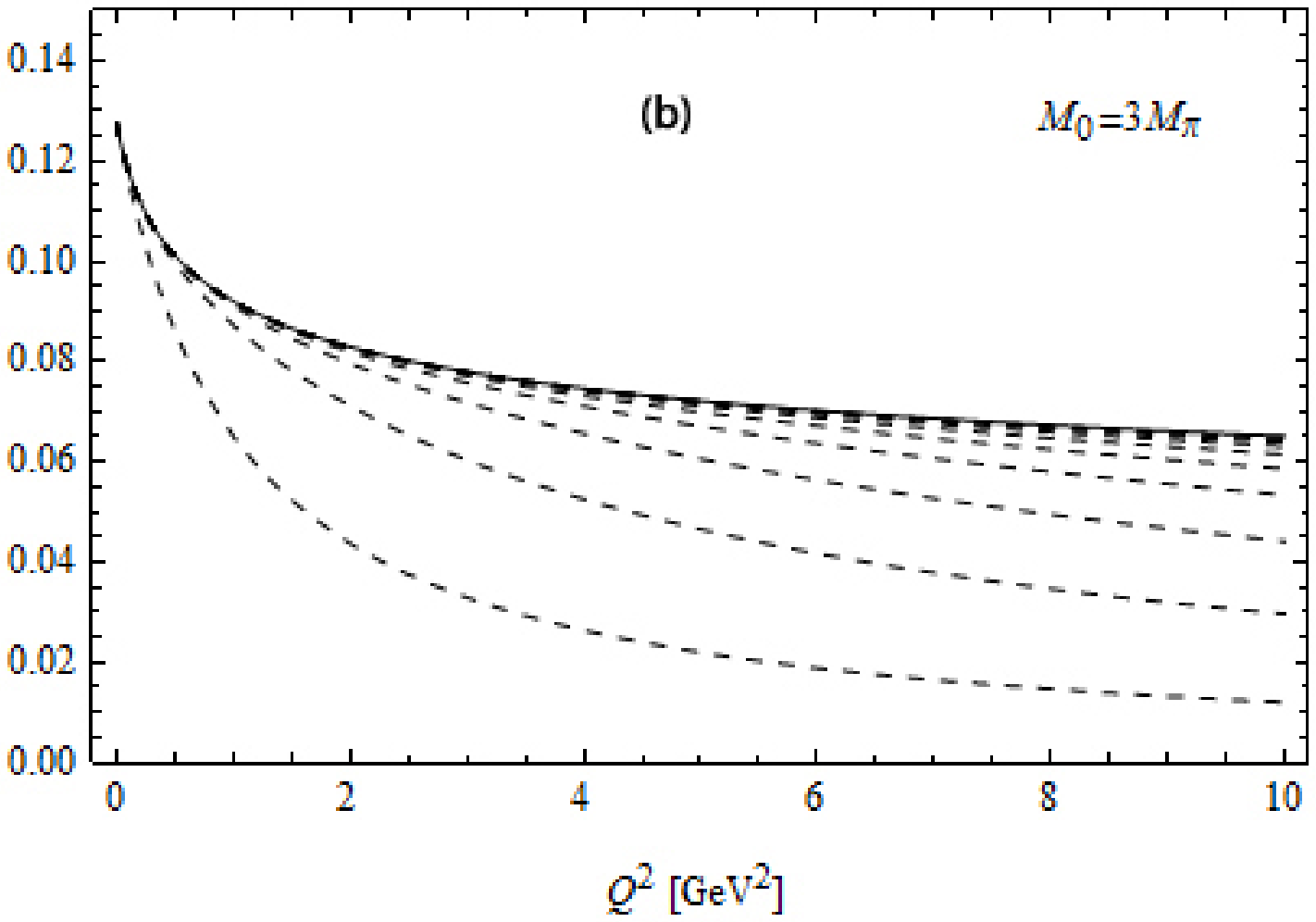,width=\linewidth}
\end{minipage}
\vspace{-0.2cm}
\caption{The full line is the ``exact'' mMA coupling, Eq.~(\ref{mMAdisp}).  
In both Figures, the dashed lines (starting from the
bottom) are all Pad\'e approximants $R^{M-1}_{M}$: 
$R^{0}_{1}$, $R^{1}_{2}$,..., $R^{9}_{10}$. The cut value $M_0$
is: (a) $M_0=2M_{\pi}$; (b) $M_0=3M_{\pi}$.}
\label{figure02ab}
\end{figure}

We set $\Lambda=0.35$ GeV (at $n_f$=3).\footnote{
In Ref.~\cite{Sh}, the value $\Lambda_{(n_f=3)} \approx 0.4$ GeV was obtained 
in MA by requiring that the MA model reproduce measured values of 
QCD observables at higher energies ($|Q| \agt 10$ GeV)
in ${\overline {\rm MS}}$ scheme.} 
This value can be
changed later when we fit experimental data. Here we will
compare numerically the accuracy of the rational approximants with
the ``exact'' numbers, i.e., those obtained by evaluating
numerically integrals Eq.~(\ref{mMAdisp}). 

In Fig.~\ref{figure01} we compare mMA
coupling (for various values of the IR cutoff $M_0$)
with the standard MA coupling\footnote{
${\mathcal{A}}_{1}^{(MA)}(0) = 1/\beta_0 \approx 0.44$; however, the
slope $d {\mathcal{A}}_1^{(MA)}(Q^2)/d Q^2$ at $Q^2=0$ is infinite.}
at low positive $Q^2$'s.
These values are calculated by performing integrals in the dispersive 
relations (\ref{mMAdisp}) and (\ref{MAdisp}) for each $Q^2$. 
An effect of increasing the value of the cutoff $M_0^2$
in Eq.~(\ref{mMAdisp}) is to decrease the values of
${\mathcal{A}}_{1}^{(mMA)}(Q^2)$ at low $Q^2$.

We can compute $L_n$ coefficients of mMA appearing in 
Eq.~(\ref{A1expu}) by using the
discontinuity function (\ref{rho12l}) in integrals (\ref{Ln}). All the
integrations are done numerically. The first twenty coefficients 
are shown in Table \ref{table1}.
With these coefficients we can compute the paradiagonal Pad{\'e} approximants
$R_{M}^{M-1}(u)$ up to $R_{10}^{9}(u)$ (for higher Pad{\'e}'s we need
more coefficients). 

All the rational approximants $R_{M}^{M-1}(u)$
up to $R_{10}^{9}(u)$ for ${\mathcal{A}}_{1}^{(mMA)}(u\Lambda^2)$ 
are shown in the Figs.~\ref{figure02ab}, for two choices of the
IR cutoff in Eq.~(\ref{mMAdisp}): $M_0 = 2 M_{\pi}$, $3 M_{\pi}$.
We recall that $M_0^2$ is the convergence radius of the 
Stieltjes function ${\mathcal{A}}_{1}^{(mMA)}(Q^2)$.
Since this radius is larger in Fig.~\ref{figure02ab}(b),
the convergence of the Pad\'e approximants is also faster there.
The rational function $R_{10}^{9}(Q^2)$ matches 
${\mathcal{A}}_{1}^{(mMA)}(Q^2)$ in Fig.~\ref{figure02ab}(a) quite well 
for all $0 < Q^2<10 \ {\rm GeV}^2$, and in
Fig.~\ref{figure02ab}(b) the agreement is even better.

In order to give an idea of how the
approximants work, we list in Table \ref{table2} the values of $Q^2$
at which the deviation from ${\mathcal{A}}_{1}^{(mMA)}(Q^2)$ 
reaches 0.1\% (the deviation increases with increasing $Q^2$),
for the case $M_0=2 M_{\pi}$. 
For the Pad{\'e} approximant $R_{M}^{M-1}(u)$ with $M=10$ the deviation
is less than 0.1\% for all $Q^2 < 2.5 \ {\rm GeV}^2$,
and less than 1\% for all $Q^2 \leq 5.4 \ {\rm GeV}^2$
(when the Pad\'e index is $M=20$, these values increase to 
$10.33 \ {\rm GeV}^2$ and $22.63  \ {\rm GeV}^2$, respectively).  
We recall that these values are much higher than the convergence
radius of the Taylor series, Eqs.~(\ref{A1expQ}) and (\ref{A1expu}):
$R= Q^2=M_0^2=0.0784 \ {\rm GeV}^2$.
We see that the first pole ($-s_1$) converges from below to the IR cut 
$-s_0=-M_0^2/\Lambda^2=-4M_{\pi}^2/\Lambda^2$
($-s_0=-0.64$ in our case) when index $M$ increases, as expected.
For example, when $M=10$, we obtain: 
$s_1 = 0.6492$ ($\approx s_0=0.6400$),
$s_2=0.6907$, etc.
When the cut $M_0$ is increased from $2 M_{\pi}$ to 
$3 M_{\pi}$ ($\Rightarrow \ -s_0=-1.44$), we obtain for $M=10$:
$s_1 = 1.4605$   ($\approx s_0=1.44$), $s_2=1.5525$, etc.
\begin{table}
\caption{\label{table1}
The first twenty
coefficients computed with the discontinuity function at two-loop
given in Eq.~(\ref{rho12l}), for $M_0=2 M_{\pi}$. In parentheses,
the results for $M_0=3 M_{\pi}$ are given. All the numbers here are presented
as rounded off at the sixth digit.}
\begin{ruledtabular}
\begin{tabular}{cccccccc}
$L_n$ && $L_n$ &&\\
\hline
$L_0$& 0.151418 (0.127410)      &$L_{10}$ & 0.270477 ($6.95492 \cdot 10^{-5}$) \\
$L_1$& -0.0408795 (-0.0149290)  &$L_{11}$ & -0.384736 ($-4.40027 \cdot 10^{-5}$) \\
$L_2$& 0.0353764  (0.00582799)  &$L_{12}$ & 0.551687 ($2.80615 \cdot 10^{-5}$) \\
$L_3$& -0.0380741 (-0.00281121) &$L_{13}$ & -0.796468 ($-1.80155 \cdot 10^{-5}$) \\
$L_4$& 0.0453111 (0.00149464)   & $L_{14}$ & 1.15653 ($1.16323 \cdot 10^{-5}$)  \\
$L_5$& -0.0571406 ($-8.40628 \cdot 10^{-4}$) & $L_{15}$ & -1.68780 ($-7.54792 \cdot 10^{-6}$)   \\
$L_6$& 0.0748244 ($4.90454 \cdot 10^{-4}$)  & $L_{16}$ &2.47387 ($4.91884 \cdot 10^{-6}$) \\
$L_7$& -0.100605 ($-2.93630 \cdot 10^{-4}$) & $L_{17}$ & -3.63999 ($-3.21771 \cdot 10^{-6}$) \\
$L_8$& 0.137943 ($1.79193 \cdot 10^{-4}$)   & $L_{18}$ & 5.37405 ($2.11200 \cdot 10^{-6}$)  \\
$L_9$& -0.192008 ($-1.10983 \cdot 10^{-4}$) & $L_{19}$ & -7.95836 ($-1.39043 \cdot 10^{-6}$)  \\
\end{tabular}
\end{ruledtabular}
\end{table}
\begin{table}
\caption{\label{table2}The values of $Q^2 = u \Lambda^2$ at which the Pad\'e 
approximants $R_{M}^{M-1}$ (up to  $R_{10}^{9}$), for $M_0=2 M_{\pi}$,
reach 0.1\% deviation from ${\mathcal{A}}_{1}^{(mMA)}$. Further, the first few 
values $s_n \equiv M_n^2/\Lambda^2$ and $f_n^2$ are shown.} 
\begin{ruledtabular}
\begin{tabular}{cccccccc}
$R_M^{M-1}(u)$ & $Q^2$ $[{\rm GeV}^2]$ & $s_1$ & $f_1^2$ & $s_2$ & $f_2^2$ & $s_3$ & $f_3^2$\\
\hline
$R_{1}^{0}$& 0.01 & 3.704 & 0.56085 & & & &\\
$R_{2}^{1}$& 0.08 & 0.912073 & 0.0263116 & 13.2502 & 1.62407 & &\\
$R_{3}^{2}$& 0.20 & 0.747835 & 0.00947914 & 1.62259  & 0.0538095 & 30.2372  & 3.19243\\
$R_{4}^{3}$& 0.38 & 0.698643 & 0.00498205 & 1.04773 & 0.0185788 & 2.62744 & 0.0831117\\
$R_{5}^{4}$& 0.60 & 0.677037 & 0.00309638 & 0.871413 & 0.00968867 & 1.46119 & 0.0275631 \\
$R_{6}^{5}$& 0.88 & 0.665567 & 0.00211868 & 0.7913 & 0.00604702 & 1.11052 & 0.0140762\\
$R_{7}^{6}$& 1.20 & 0.658731 & 0.00154383 & 0.747386 & 0.00417231  & 0.951402 & 0.00869082\\
$R_{8}^{7}$& 1.58 & 0.654321 & 0.00117625 & 0.720466 & 0.00306913 & 0.863709 & 0.00596662\\
$R_{9}^{8}$& 2.02 & 0.651309  & 0.00092658 & 0.702677 & 0.00236037 & 0.809515 & 0.00438199\\
$R_{10}^{9}$&2.50 & 0.649158 & 0.000749085 & 0.690269 & 0.00187583 &  0.773381 & 0.0033713\\
\end{tabular}
\end{ruledtabular}
\end{table}

\section{Approximants for the higher power analogs of the analytic coupling}
\label{sec:AnPA}

To obtain the analogs of the higher powers of the analytic coupling
in this formalism, we use relations given in Ref.
\cite{CV1}. At the 3-loop level truncated series for the
analytic coupling, we have:
\begin{eqnarray} \label{A2A3}
{\mathcal{A}}_{2}(Q^2)&=&\
\tilde{{\mathcal{A}}_{2}}(Q^2)-c_{1}\tilde{{\mathcal{A}}}_{3}(Q^2),
\qquad
{\mathcal{A}}_{3}(Q^2)=\tilde{{\mathcal{A}}_{3}}(Q^2),
\end{eqnarray} 
where
\begin{equation} \label{tAn}
\tilde{{\mathcal{A}}_n}(Q^2)=
\frac{(-1)^{n-1}}{\beta_{0}^{n-1}(n-1)!}
\frac{\partial^{n-1}{\mathcal{A}}_1(Q^2)}{\partial(\ln Q^2)^{n-1}}.
\end{equation} 
The correspondence between the powers $a^k$ of the perturbative
coupling $a(Q^2) = \alpha_s(Q^2)/\pi$ and the above quantities is:
$a^k\mapsto {\mathcal{A}}_k$. 
The couplings ${\mathcal{A}}_k$ are
the analytic versions (``analogs'')
of higher powers needed for evaluation of observables. 
We note that in general ${\mathcal{A}}_k \neq {\mathcal{A}}_1^k$
(for further discussion, c.f.~Sec.~III of Ref.~\cite{CV1}).

Using the dispersive integral expression Eq.~(\ref{mMAdisp}) in Eqs.~(\ref{tAn}), 
we obtain explicit expressions for ${\mathcal{A}}_k$'s
in terms of integrals of the (perturbative) discontinuity function $\rho_1$
\begin{eqnarray}
{\mathcal{A}}_{2}^{(mMA)}(u\Lambda^2)&=&
\frac{u}{\beta_0\pi}
\int_{s_0}^{\infty}\frac{\rho_1(s\Lambda^2)}
{(s+u)^2}ds-c_1{\mathcal{A}}_{3}^{(mMA)}(u\Lambda^2),
\label{A2} 
\\
{\mathcal{A}}_{3}^{(mMA)}(u\Lambda^2)&=& 
\frac{-u}{2\beta_0^2\pi}\int_{s_0}^{\infty}
\frac{\rho_1(s\Lambda^2)(s-u)}{(s+u)^3}ds.
\label{A3}
\end{eqnarray}
We see from here that the quantities ${\mathcal{A}}_k^{(mMA)}(Q^2)$ 
have the same location of the singularities as 
${\mathcal{A}}_{1}^{(mMA)}$: $Q^2<-M_0^2$ $(=-\Lambda^2 s_0)$. 
On the other hand, the poles $Q^2=-M_n^2 = -s_n \Lambda^2$ 
($n=1,\ldots,M$) of the Pad{\'e} approximants $R_M^{M-1}$
for ${\mathcal{A}}_1$, Eq.~(\ref{A1R}), do reflect, in a discretized
manner, the singularity cut of ${\mathcal{A}}_{1}^{(mMA)}$;
namely, these poles appear on the negative real axis in the range:
$-\infty < - M^2_M < -M^2_{M-1} < \cdots < - M^2_1$ ($< M^2_0$).
Such a pole structure of $R_M^{M-1}(u)$
is guaranteed because ${\mathcal{A}}_{1}^{(mMA)}$ is a
Stieltjes function \cite{Pade1}. On the other hand, the analytic
higher power analogs ${\mathcal{A}}_k^{(mMA)}(Q^2)$ ($k \geq 2$)
are no longer Stieltjes functions. Therefore, if we construct the Pad\'e 
approximants $R_M^{M-1}(u)$ for ${\mathcal{A}}_k^{(mMA)}(Q^2)$ ($k \geq 2$)
in the usual way, i.e., based on the first $2 M$ coefficients
of their Taylor expansion around $Q^2=0$,
there is no longer the guarantee that all the poles of 
the obtained rational functios lie on the negative real axis
(and below $-M_0^2$).
For instance, the Pad{\'e} approximant $R_{10}^{9}$ computed from 
the series of ${\mathcal{A}}_3$ has a non-physical pole on the 
positive real axis. One way to avoid this problem is to compute 
the rational approximants of ${\mathcal{A}}_k$'s directly from
the derivatives of the Pad{\'e} approximant of ${\mathcal{A}}_1^{(mMA)}$. 
If $R_{M}^{M-1}(u)$ are the paradiagonal Pad\'es that approximate
${\mathcal{A}}_1^{(mMA)}(u \Lambda^2)$, we shall define the rational approximants 
$R_2(u)$ and $R_3(u)$ for ${\mathcal{A}}_2^{(mMA)}(u \Lambda^2)$ and 
${\mathcal{A}}_3^{(mMA)}(u \Lambda^2)$, respectively, in the following way:
\begin{eqnarray} \label{R2R3}
R_{2}(u)&=&\tilde{R_{2}}(u)-c_{1}\tilde{R}_{3}(u),
\qquad R_{3}(u)=\tilde{R_{3}}(u),
\end{eqnarray} 
where
\begin{equation} \label{tRn}
\tilde{R_n}(u)=\frac{(-1)^{n-1}}{\beta_{0}^{n-1}(n-1)!}
\frac{\partial^{n-1}R_{M}^{M-1}(u)}{\partial(\ln
u)^{n-1}}.
\end{equation}
\begin{figure}[htb]
\begin{minipage}[b]{.49\linewidth}
\centering\epsfig{file=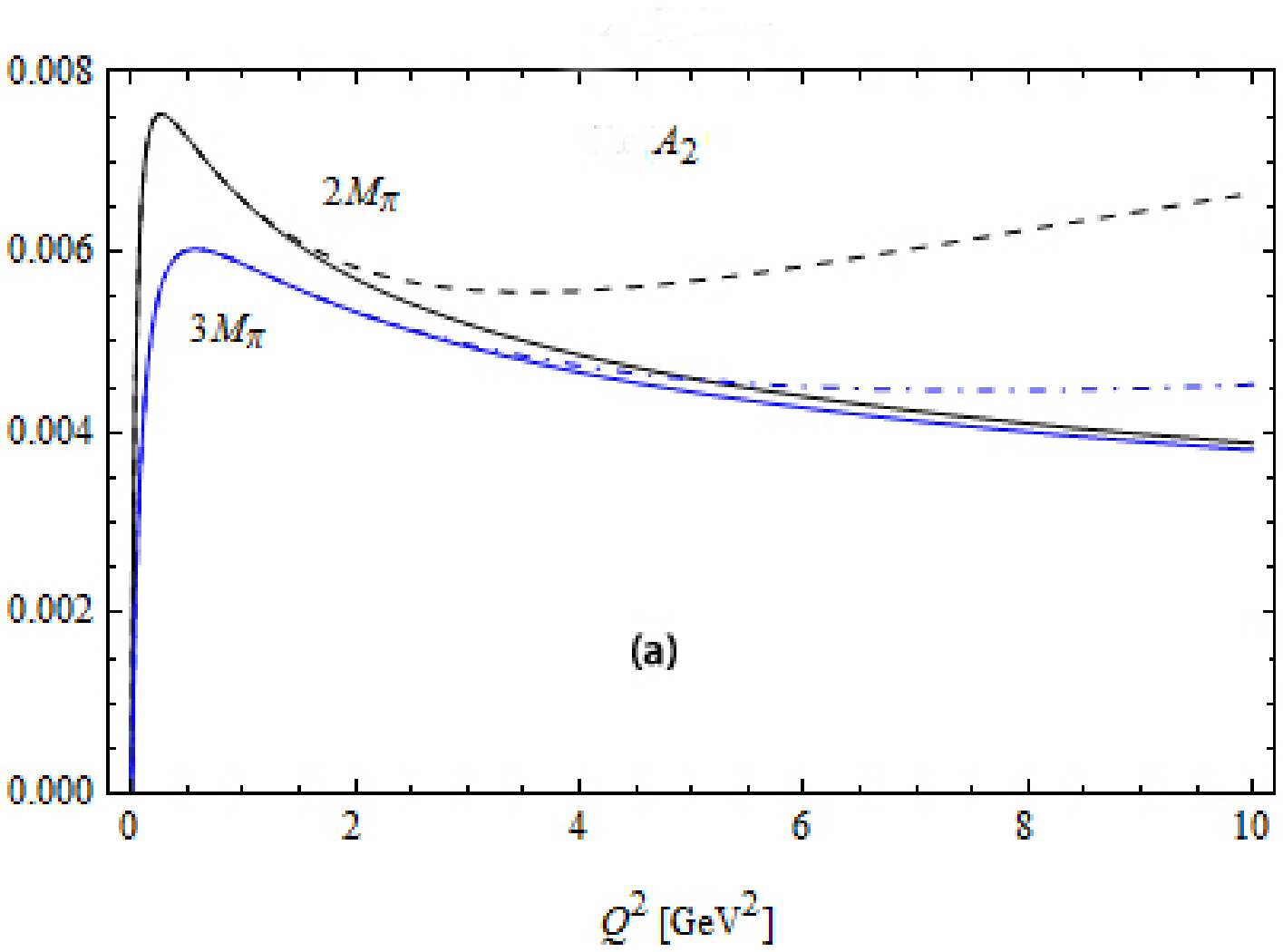,width=\linewidth}
\end{minipage}
\begin{minipage}[b]{.49\linewidth}
\centering\epsfig{file=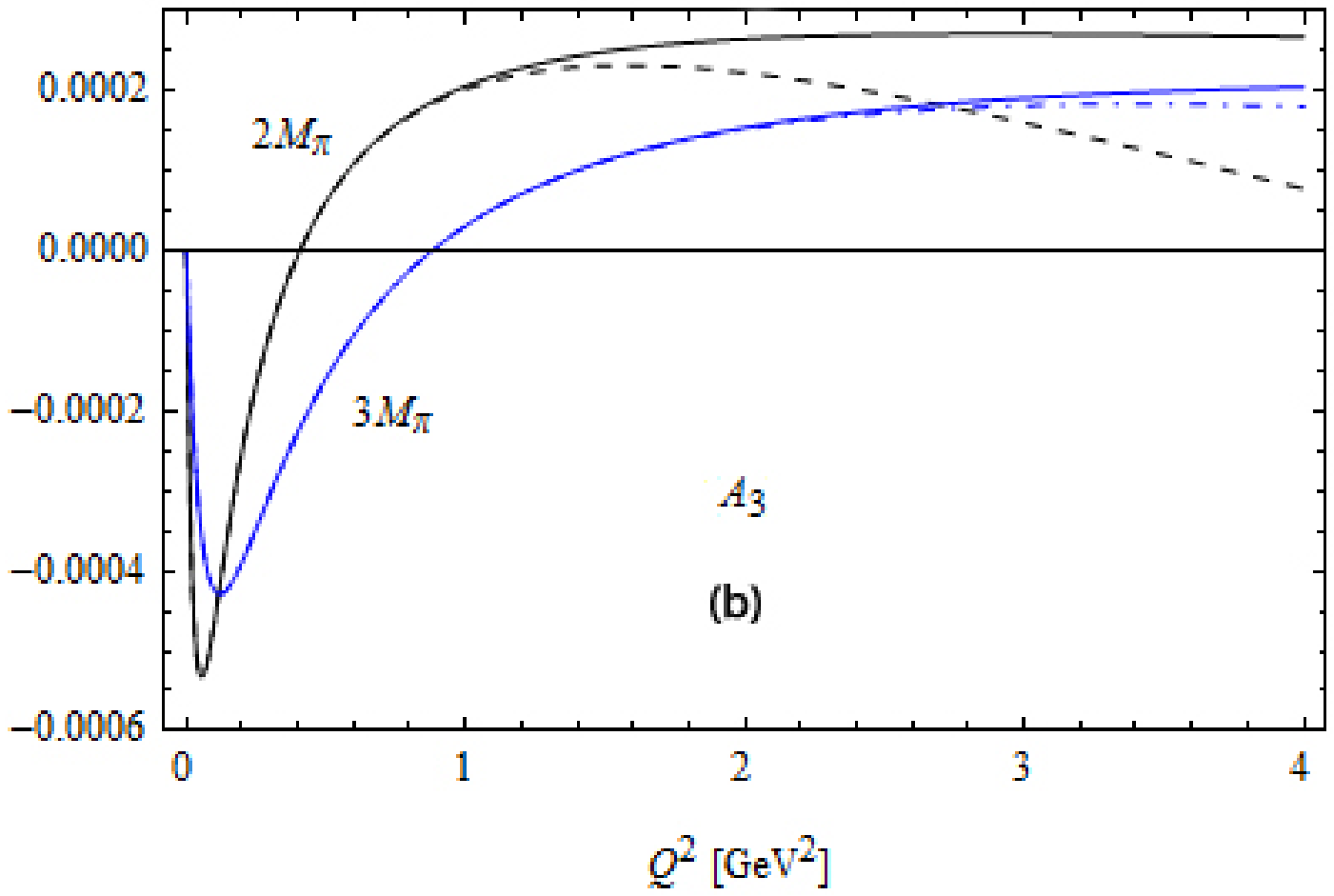,width=\linewidth}
\end{minipage}
\vspace{-0.2cm}
\caption{Comparison between: (a) ${\mathcal{A}}_2^{(mMA)}$ (continuous) 
and $R_2$ (dashed and dot-dashed), for two choices of IR cut $M_0=2 M_{\pi}$
and $M_0=3 M_{\pi}$, respectively; (b) the same as in (a), but now for 
${\mathcal{A}}_3^{(mMA)}$ and  $R_3$. Pad\'e index is $M=10$.}
\label{figure03ab}
\end{figure}
A comparison between $R_{k}$ and their respective
${\mathcal{A}}_k$, computed from $R_{10}^{9}(u)$ (i.e., Pad\'e index $M=10$),
is shown in Figs.~\ref{figure03ab}. The accuracy of these 
approximants decreases notably compared to the $R_{M}^{M-1}$
approximants of ${\mathcal{A}}_{1}^{(mMA)}$.
The accuracy of the approximants increases when the cut $M_0$ increases,
e.g., from $2 M_{\pi}$ to $3 M_{\pi}$.

\section{Complex arguments}
\label{sec:CAs}

\begin{figure}[htb]
\begin{minipage}[b]{.49\linewidth}
\centering\epsfig{file=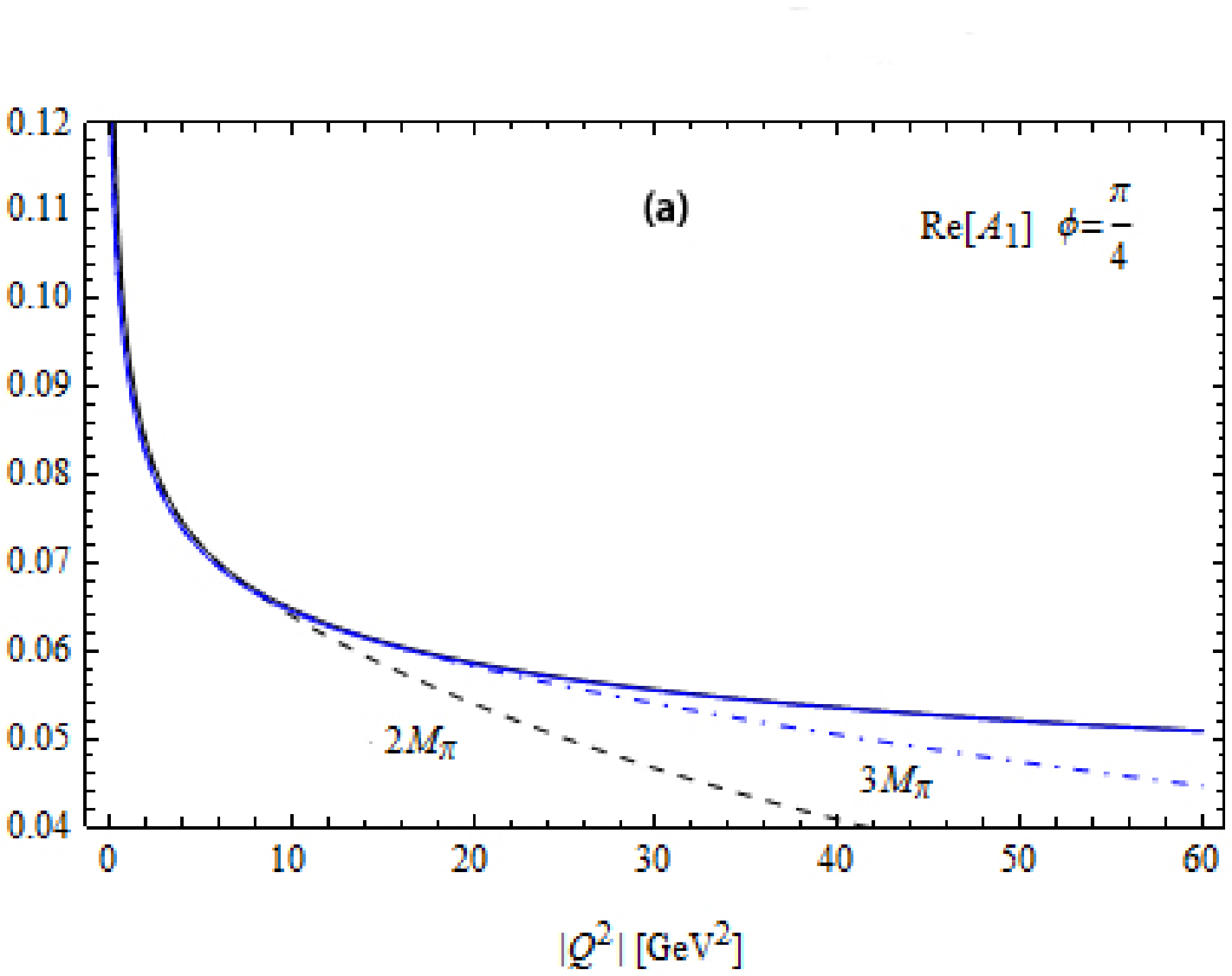,width=\linewidth}
\end{minipage}
\begin{minipage}[b]{.49\linewidth}
\centering\epsfig{file=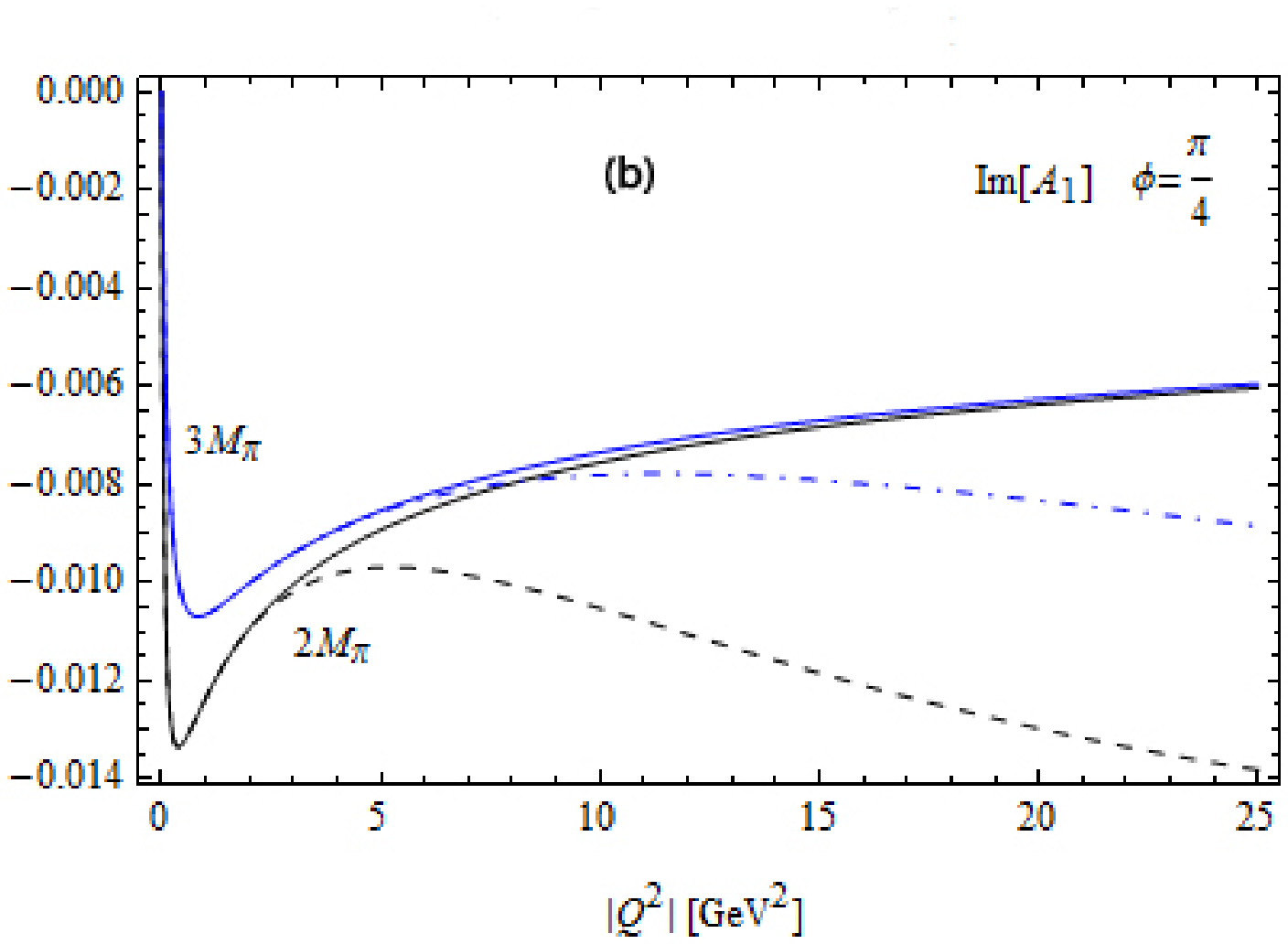,width=\linewidth}
\end{minipage}
\vspace{-0.2cm}
\caption{(a) Real parts of ${\mathcal{A}}_1^{(mMA)}$ and $R_{10}^{9}$ 
with complex arguments  $Q^2=|Q^2| \exp(i \pi/4)$; (b) same as in (a),
but for imaginary parts.}
\label{figure04ab}
\end{figure}

\begin{figure}[htb]
\begin{minipage}[b]{.49\linewidth}
\centering\epsfig{file=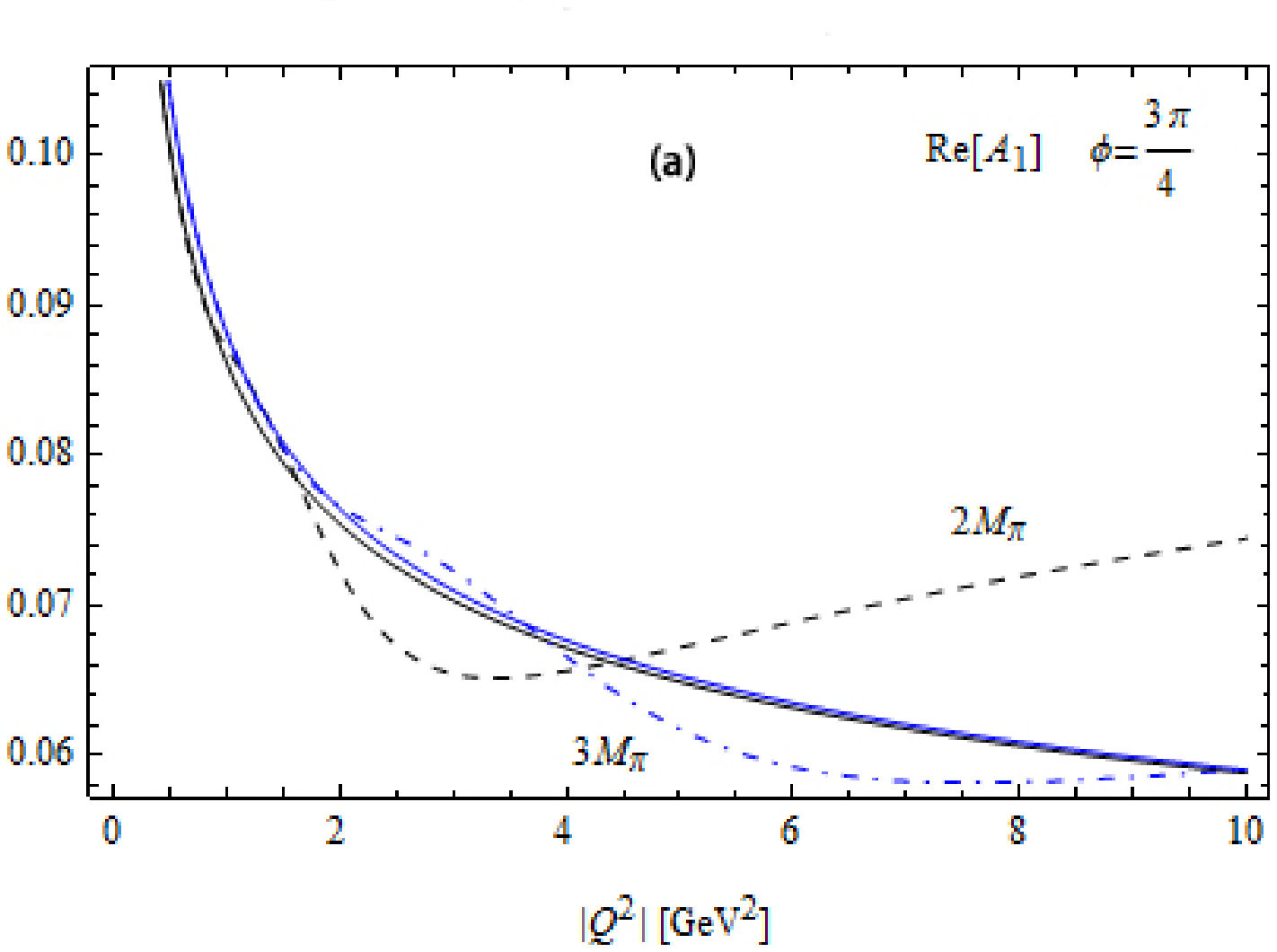,width=\linewidth}
\end{minipage}
\begin{minipage}[b]{.49\linewidth}
\centering\epsfig{file=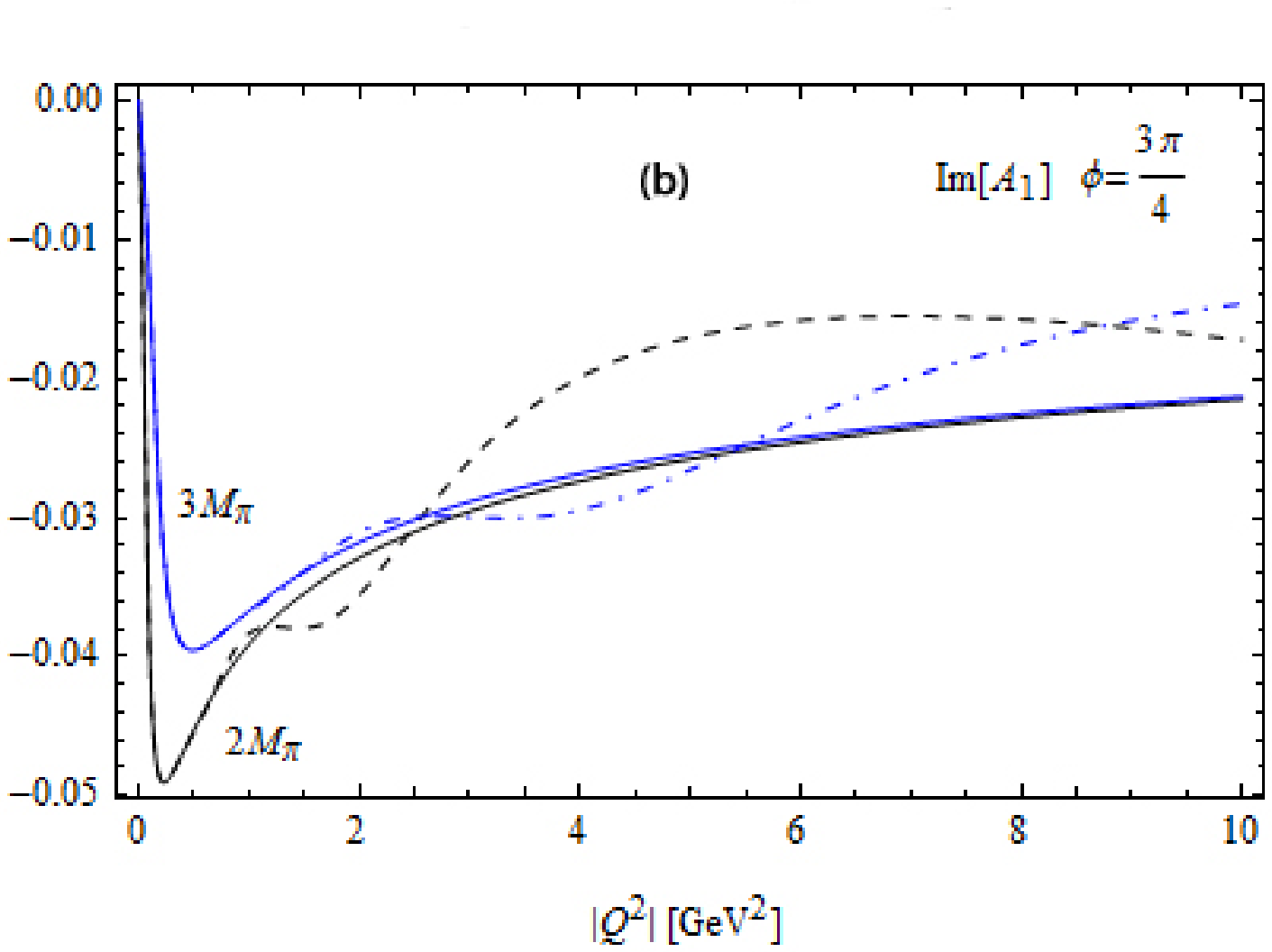,width=\linewidth}
\end{minipage}
\vspace{-0.2cm}
\caption{(a) Real parts of ${\mathcal{A}}_1^{(mMA)}$ and $R_{10}^{9}$ 
with complex arguments $Q^2=|Q^2| \exp(i 3 \pi/4)$; (b) same as in (a),
but for imaginary parts.}
\label{figure05ab}
\end{figure}

\begin{figure}[htb]
\begin{minipage}[b]{.49\linewidth}
\centering\epsfig{file=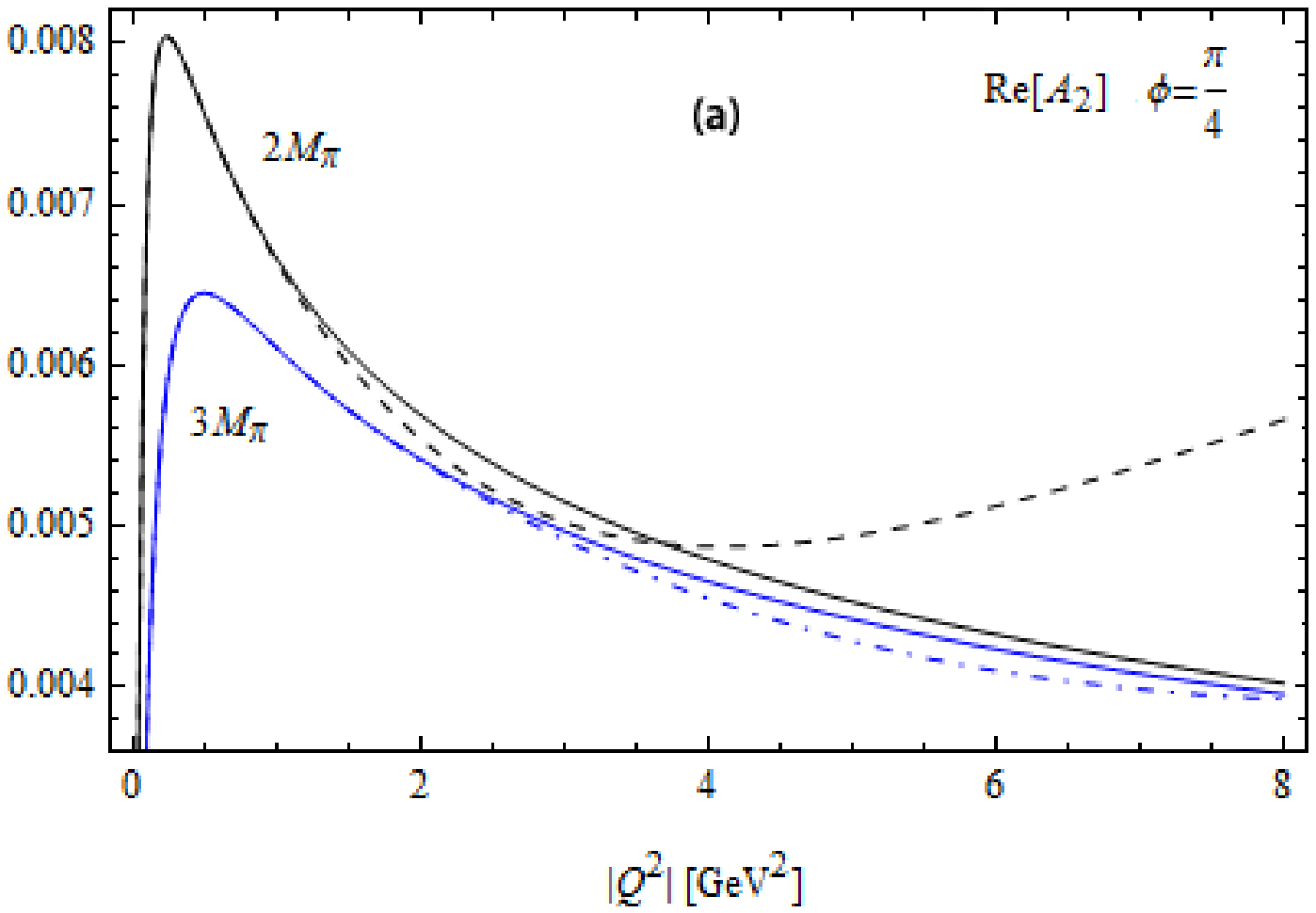,width=\linewidth}
\end{minipage}
\begin{minipage}[b]{.49\linewidth}
\centering\epsfig{file=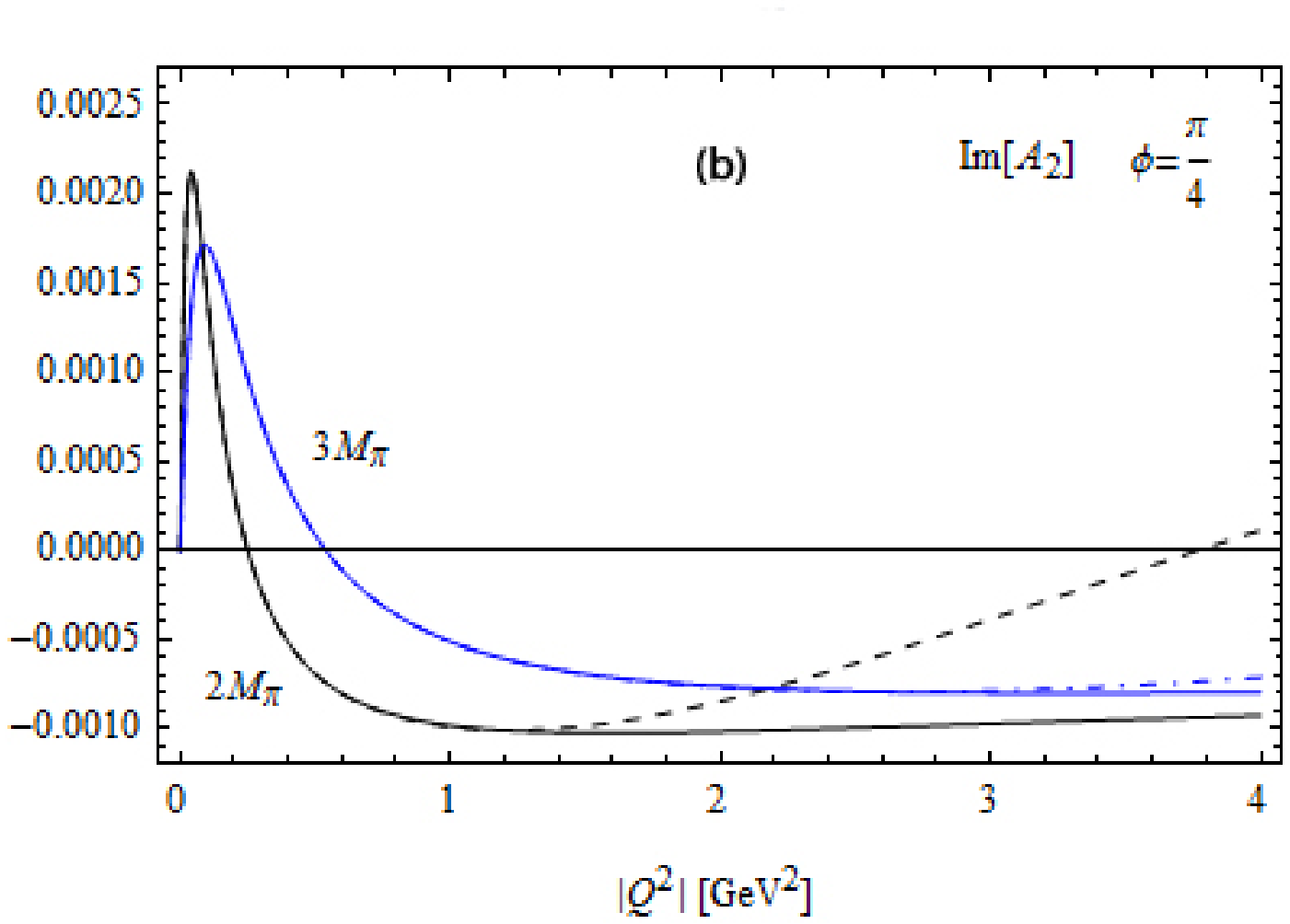,width=\linewidth}
\end{minipage}
\vspace{-0.2cm}
\caption{(a) Real parts of ${\mathcal{A}}_2^{(mMA)}$ and $R_2$ 
with complex arguments $Q^2=|Q^2| \exp(i \pi/4)$; (b) same as in (a),
but for imaginary parts.}
\label{figure06ab}
\end{figure}

\begin{figure}[htb]
\begin{minipage}[b]{.49\linewidth}
\centering\epsfig{file=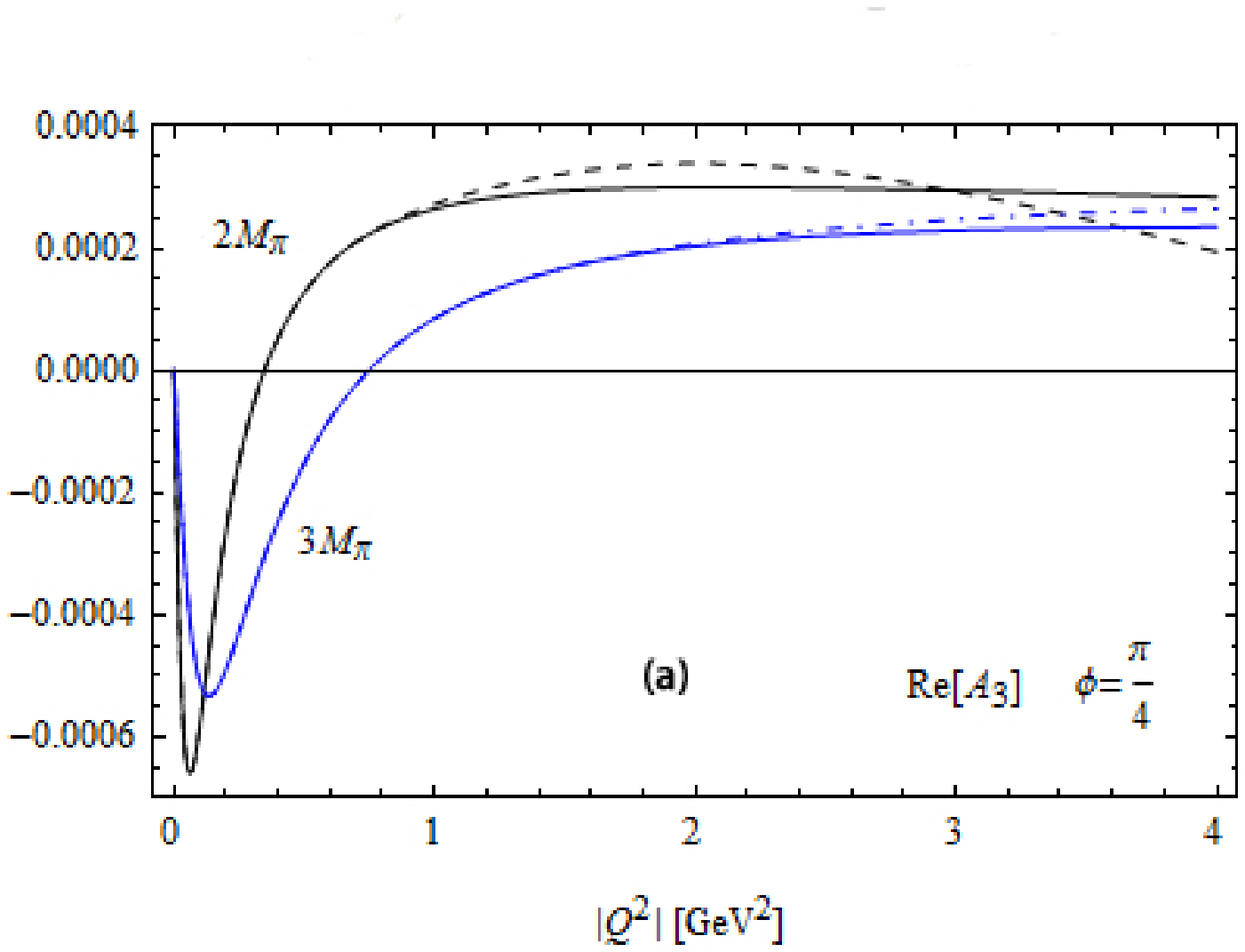,width=\linewidth}
\end{minipage}
\begin{minipage}[b]{.49\linewidth}
\centering\epsfig{file=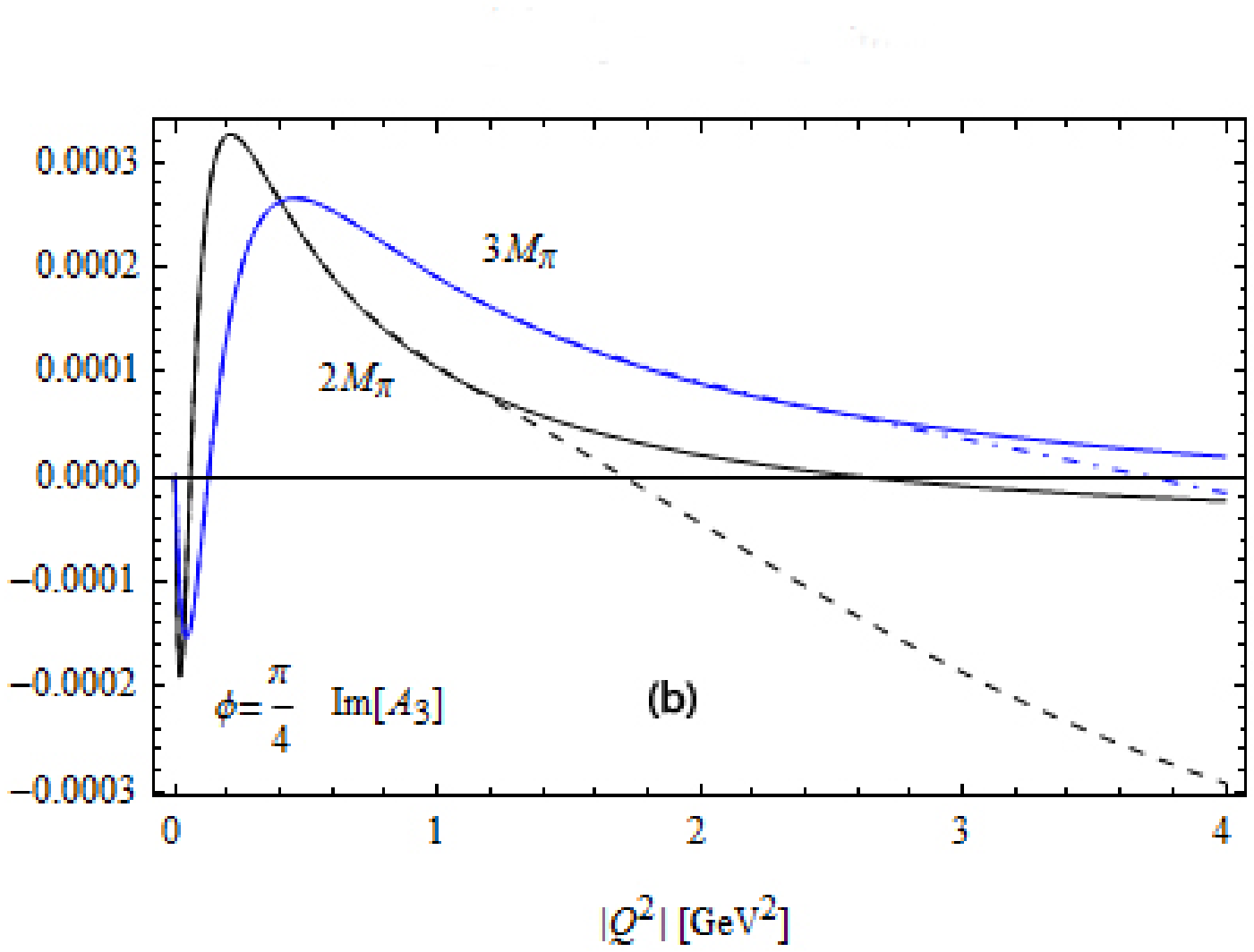,width=\linewidth}
\end{minipage}
\vspace{-0.2cm}
\caption{(a) Real parts of ${\mathcal{A}}_3^{(mMA)}$ and $R_3$ 
with complex arguments $Q^2=|Q^2| \exp(i \pi/4)$; (b) same as in (a),
but for imaginary parts.}
\label{figure07ab}
\end{figure}

In evaluation of observables, sometimes evaluation of the analytic
coupling and their power analogs at complex values of arguments 
is needed (e.g., see App.~C of Ref.~\cite{CV1}). 
For some complex arguments $Q^2=|Q^2| \exp(i \phi)$
[$u \equiv Q^2/\Lambda^2 = |u|  \exp(i \phi)$], the results
based on the approximant $R_{10}^9(u)$ (Pad\'e index $M=10$), 
are shown in Figs.~\ref{figure04ab}-\ref{figure07ab}. 
When $\phi=\pi/4$, Figs.~\ref{figure04ab} (a), (b) show that
the approximant for ${\mathcal{A}}_{1}^{(mMA)}(Q^2)$ 
works with less than 1\% error up to $|Q^2| \approx 10 \ {\rm GeV}^2$ 
for the real part, and $|Q^2| \approx 3 \ {\rm GeV}^2$ for
the imaginary part. The approximants $R_2(u)$ and $R_3(u)$
for ${\mathcal{A}}_2$ and ${\mathcal{A}}_3$,
(see Figs.~\ref{figure06ab}-\ref{figure07ab})
fail already at lower $|Q^2|$. 
Figs.~\ref{figure05ab} (a), (b) show the approximant
for ${\mathcal{A}}_{1}^{(mMA)}(Q^2)$
when $\phi= 3 \pi/4$.   
Figs.~\ref{figure04ab}-\ref{figure07ab} indicate that the relative 
accuracy of our approximants decreases: (I) when the index $n$ of the 
power analog ${\mathcal{A}}_n$ increases; (II) when the considered ray 
$Q^2=|Q^2| \exp(i \phi)$ comes closer to the time-like semiaxis ($|\phi|$ 
closer to $\pi$). Aspect (I) can be understood from the fact that, 
by our construction, $R_n$ involves $(n-1)$ derivatives of the Pad\'e 
approximant $R_1\equiv R_M^{M-1}$. Aspect (II) is also to be expected, 
because rays with $\phi \approx \pm\pi$ are close to the singularity cut 
of ${\mathcal{A}}_n$'s.

\begin{figure}[htb]
\begin{minipage}[b]{.49\linewidth}
\centering\epsfig{file=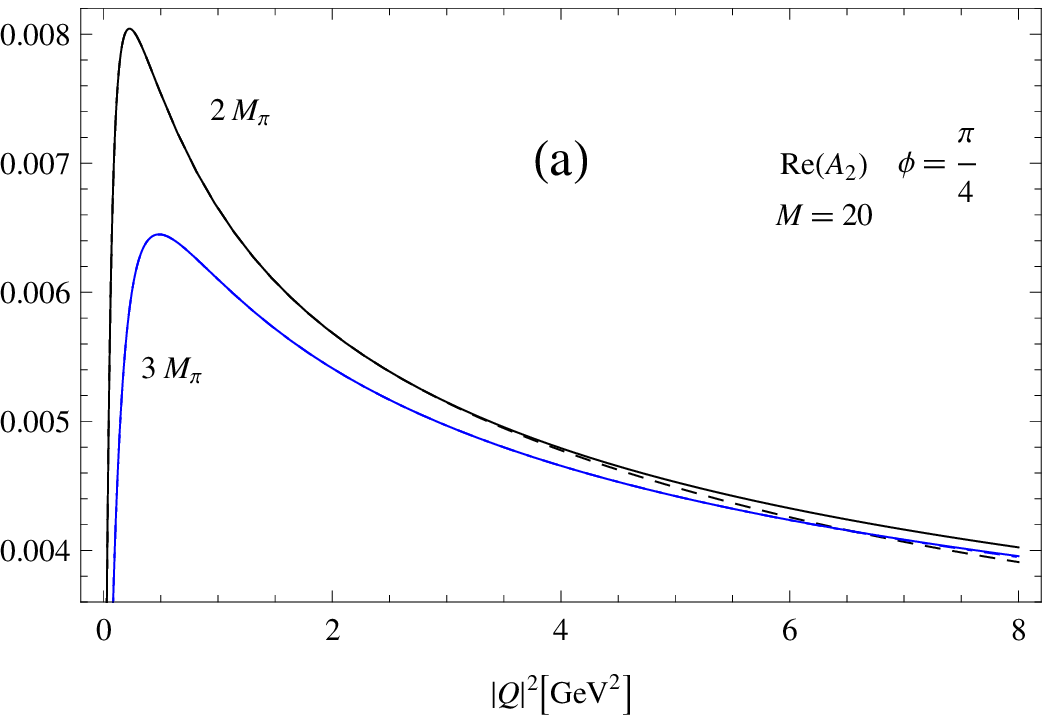,width=\linewidth}
\end{minipage}
\begin{minipage}[b]{.49\linewidth}
\centering\epsfig{file=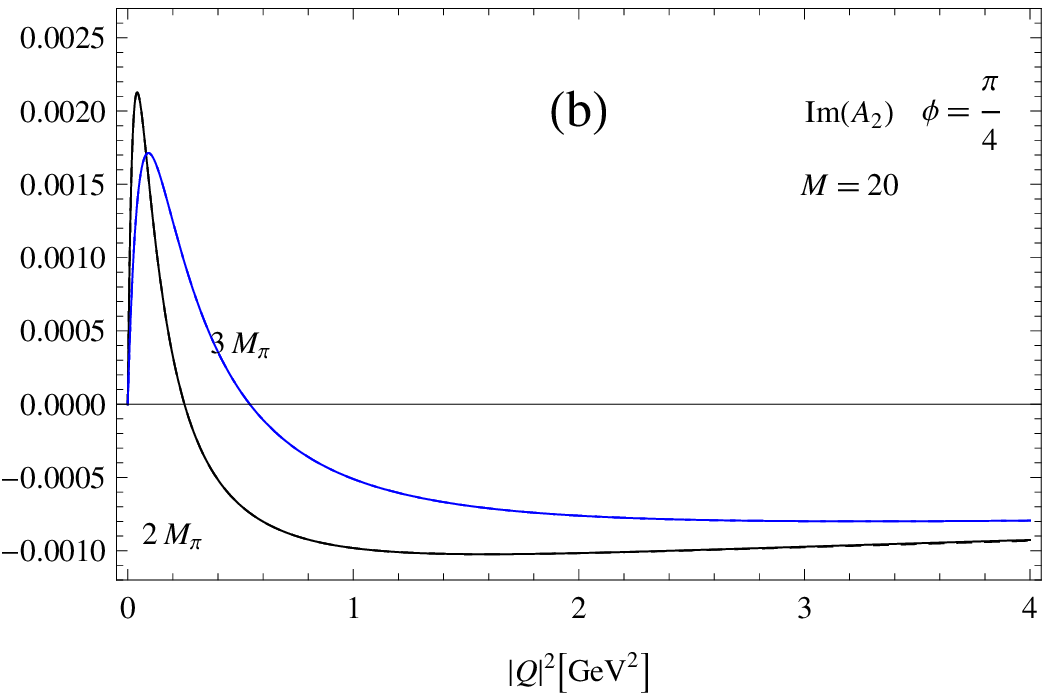,width=\linewidth}
\end{minipage}
\vspace{-0.2cm}
\caption{Now $R_2$ is computed 
from the Pad\'e $R_{20}^{19}$ of ${\mathcal{A}}_1^{(mMA)}$ ($M=20$):
(a) real parts of ${\mathcal{A}}_2^{(mMA)}$ (continuous) and $R_2$ (dashed and
dot-dashed), for the IR cut value $M_0=2 M_{\pi}$, $3 M_{\pi}$, respectively,
with complex arguments $Q^2=|Q^2| \exp(i \pi/4)$; 
(b) same as in (a), but for imaginary parts. For $M_0=3 M_{\pi}$, no
deviations of $R_2$ from ${\mathcal{A}}_2^{(mMA)}$ can be seen by the eye.}
\label{figure08ab}
\end{figure}

The accuracy can be increased if we compute 
$R_1 \equiv  R_M^{M-1}$ as a higher order Pad\'e (higher index $M$). 
For instance, Figs.~\ref{figure08ab} (a), (b)
show the approximants for ${\mathcal{A}}_2^{(mMA)}$, with $\phi=\pi/4$, 
which are calculated from $R_{20}^{19}(u)$ (i.e., Pad\'e index $M=20$). 
The latter is calculated from the first 40 coefficients of the Taylor
series of ${\mathcal{A}}_1^{(mMA)}(u\Lambda^2)$. We see that
in the case of $M_0=2 M_{\pi}$, the deviation of ${\rm Re}[R_2]$ from 
${\rm Re}[{\mathcal{A}}_2]$ becomes discernible to the eye (1\% deviation) only for
$Q^2 > 5 \ {\rm GeV}^2$; this is to be compared with Fig.~\ref{figure06ab} (a)
(where Pad\'e index $M=10$).
 If we used $\phi=0$ instead (see Figs.~\ref{figure03ab}
where $M=10$), the deviations of  ${\rm Re}[R_2]$ from 
${\rm Re}[{\mathcal{A}}_2]$, for $M=20$ and $M_0=2 M_{\pi}$, 
on the scale of Figs.~\ref{figure08ab}
would turn out to be discernible to the eye only for 
$Q^2 > 6 \ {\rm GeV}^2$, i.e., the convergence turns 
out to be even better than in the case $\phi=\pi/4$.
The deviations of ${\rm Im}[R_2]$ from ${\rm Im}[{\mathcal{A}}_2]$
in Fig.~\ref{figure08ab} (b) are not discernible to the eye.
We also see that the deviations of $R_2$ from ${\mathcal{A}}_2$
cannot be seen by the eye in Figs.~\ref{figure08ab} when the 
cut value $M_0$ increases to $3 M_{\pi}$.\footnote{
At the level $M=20$, it is important that the 40 Taylor coefficients
$L_n$ and the coefficients of the Pad\'e approximant $R_{M}^{M-1}$ be
calculated to high accuracy (to at least about 20 and 30 digits, respectively),
in order to avoid numerical instabilities connected with cancellation of
large numbers.}

Both aforementioned aspects (I, II) that decrease accuracy are, however, 
not very important in practice when evaluating observables.
Namely, the higher order 
contributions ($\sim {\mathcal{A}}_n$) are very suppressed in anQCD
(even when $|Q^2|$ is low),
and the contributions of $Q^2$ near the time-like axis in the contour-type 
of integrations (e.g., for the semihadronic $\tau$ decay ratio $r_{\tau}$) 
are usually suppressed by the integrand. 

\section{Prospects of applications in fitting experimental data}
\label{sec:expd}

In Eq.~(\ref{mMAdisp}) we considered the dispersive relation for
${\mathcal{A}}_1$ with an IR $\sigma$-cutoff $\sigma_{\rm cut}=M_0^2$
($\sim M_{\pi}^2$) imposed on the perturbative QCD discontonuity
function $\rho_1(\sigma) \equiv {\rm Im} [a(Q^2=-\sigma- i \epsilon)]$.
Such a cutoff results in the analyticity of ${\mathcal{A}}_1$
around $Q^2=0$, thus reflecting the analyticity
of space-like observables ${\cal D}(Q^2)$ in the complex
plane excluding the time-like semiaxis but including a regime
around $Q^2=0$. 

In MA model, the scale $\Lambda$ can be
fixed so that it reproduces measured values of QCD observables 
at higher energies $Q \agt 10$ GeV ($\Rightarrow \Lambda_{n_f=3} \approx
0.4$ GeV) \cite{Sh}. 
However, then MA gives too low value
of the massless strangeless  ($\triangle S=0$) semihadronic $\tau$-decay ratio:
$r_{\tau} \approx 0.14$  \cite{Milton:2000fi, CV1}.
The experimentally measured value of this quantity is:
$r_{\tau} = 0.204 \pm 0.005$ \cite{ALEPH}.
The latter value can be reproduced in MA with 
$\Lambda_{(n_f=3)} \approx 0.4$ GeV
only if the current masses of light quarks ($m_u, m_d \sim 1$ MeV)
are replaced by much larger (constituent) masses
($m_u, m_d \approx 0.25$ GeV) \cite{Milton:2001mq}
and the threshold effects become very important. 

\begin{figure}[htb]
\centering\epsfig{file=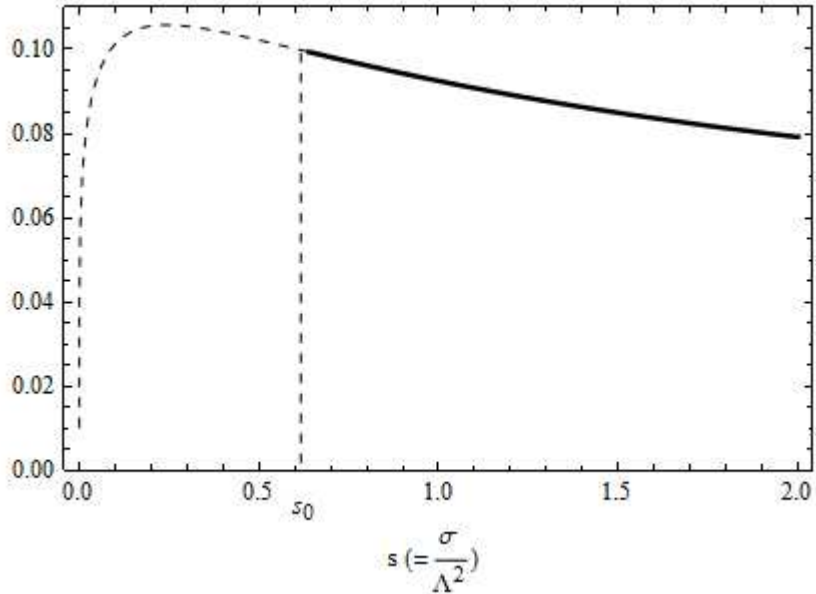,width=12.cm}
\vspace{0.cm}
\caption{This Figure shows the IR cut
$s_0=M_0^2/\Lambda^2$ (when $M_0=2 M_{\pi}$) for perturbative $\rho_1$ that
was used starting in Eq.~(\ref{mMAdisp}). The unknown low momentum
part ($0 < s \alt s_0$) can be simulated by adding a Dirac delta,
say $\pi f_{-1}^2 \delta(s - s_{-1})$, where
$0 < s_{-1} \alt s_0$.}
\vspace{-0.4cm}
\label{figure09}
\end{figure}

By introducing IR cutoff $\sigma_{\rm cut}=M_0^2$, the coupling
${\mathcal{A}}_1$ gets further diminished at low
$Q^2$, and thus further diminishes the value of $r_{\tau}$.
To remedy this situation, we can, in the simplest way,
simulate the unknown behavior of $\rho_1(\sigma)$
(Fig.~\ref{figure09})
at $\sigma \alt M_0^2$ by adding a simple positive Dirac delta peak:
$\delta \rho_1(s \Lambda^2) = \pi f_{-1}^2 \delta(s - s_{-1})$,
where $0 < s_{-1} \alt s_0$ ($\equiv M_0^2/\Lambda^2$).
This would then allow us to achieve, in the model, 
the correct value of $r_{\tau}$ while still maintaining the
analyticity of ${\mathcal{A}}_1$ around $Q^2=0$.
Thus, the full discontinity function in such a
``delta-modified'' MA model (dmMA) is
\begin{equation}
\rho_1^{\rm (dmMA)}(s \Lambda^2) = 
\Theta(s - s_0) \rho_1(s \Lambda^2)
+  \pi f_{-1}^2 \delta(s - s_{-1}) \ ,
\label{dmrho1}
\end{equation}
where $\Theta(x)$ is the Heaviside step function ($+1$ for $x>0$, zero otherwise),
and $\rho_1(s \Lambda^2)$ is the pertubative QCD discontinuity function:
$\rho_1(s \Lambda^2) = a(-s \Lambda^2 - i \epsilon)$.
This leads to the following ${\mathcal{A}}_1$:
\begin{equation}
 {\mathcal{A}}_1^{(dmMA)}(u \Lambda^2) = \frac{1}{\pi} \int_{s_0}^{\infty} ds 
\ \frac{\rho_1(s \Lambda^2)}{s + u} + \frac{f_{-1}^2}{u+s_{-1}} \ .
\label{dmA1}
\end{equation}
Applying the Pad\'e $R_M^{M-1}(u)$ approximation to this analytic coupling
we obtain
\begin{equation}
 {\mathcal{A}}_1^{\rm (dmMA)}(u \Lambda^2) \approx R_M^{M-1}(u) =
\sum_{n=-1 (n \not=0)}^M \frac{f_{n}^2}{u+s_{n}}
= \sum_{n=-1 (n \not=0)}^M \frac{F_{n}^2}{Q^2+M_{n}^2} \ .
\label{dmA1P}
\end{equation}
This has the same form as the Pad\'e $R_M^{M-1}(u)$ applied
to ${\mathcal{A}}_1^{\rm (mMA)}$ (MA with IR cut), Eq.~(\ref{A1R}),
but just with one more term ($n=-1$).
This model has three dimensionless model parameters:
$s_0, s_{-1}, f_{-1}$. All are positive and $\sim 1$.
As presented, the model is considered in the 't Hooft
scheme ($\beta_j=0$ for all $j \geq 2$). The scale parameter
$\Lambda_{n_f=3}$ is fixed by fitting the model to experimental values
of observables at high energies ($|Q| \agt 10$ GeV), such as
$\Upsilon$ decay, $e^{+}e^{-} \to$ hadrons, $Z \to$ hadrons.    
The values of low energy QCD observables ($|Q| \sim 1$ GeV), such as
$r_{\tau}$ and Bjorken polarized sum rule,
are sensitive to the values of parameters $s_0, s_{-1}$ and $f_{-1}$;
therefore, the latter are to be fixed by fitting to the
experimental values of such observables.

We see that even with this modification (dmMA) of the MA,
the evaluation of the couplings
${\mathcal{A}}_n(u \Lambda^2)$ ($n=1,2,\ldots$)
is made simple and efficient by using the Pad\'e approximant
(\ref{dmA1P}); for the evaluation, it suffices
to know the (three) parameters $s_0, s_{-1}$ and $f_{-1}$
and the first few coefficients $L_k$ [Eqs.~(\ref{A1expu})-(\ref{Ln})
and Table \ref{table1}]. 

\section{Summary}
\label{sec:summ}

We worked with the minimal analytic (MA) model modified (mMA) by
an IR cutoff $\sigma_{\rm cut} = M_0^2 \sim M_{\pi}^2$
for the perturbative discontinuity function.
In such a model, the analytic coupling ${\mathcal{A}}_1(Q^2)$
is analytic in the entire $Q^2$-complex plane excluding the
semi-axis $Q^2 < -M_0^2$. The analytic properties of such (mMA) coupling
reflect the analytic properties of space-like
QCD observables ${\cal D}(Q^2)$, among them analyticity 
in the point $Q^2=0$ and its vicinity. Further, such a (mMA) coupling
is a Stieltjes function of $Q^2$. This implies
that it will be efficiently approximated by (paradiagonal)
Pad\'e approximants $R_{M}^{M-1}(Q^2)$, i.e.,
$R_{M}^{M-1}(Q^2)$ converges to ${\mathcal{A}}_1(Q^2)$
at any point of analyticity when index $M$ increases ($M \to \infty$).
The coupling ${\mathcal{A}}_1(Q^2)$ in the form of
$R_{M}^{M-1}(Q^2)$ can be easily and efficienty evaluated 
[i.e., without performing time-consuming dispersion-type
integrations (\ref{mMAdisp}) for each $Q^2$]
just by knowing the first few coefficients of Taylor expansion
of ${\mathcal{A}}_1(Q^2)$ in powers of $Q^2$.

We showed that for real and complex arguments the paradiagonal
Pad{\'e} approximants of the analytic coupling 
${\mathcal{A}}_1^{(mMA)}(Q^2)$
are precise at low positive $Q^2$'s. This high precision
range of positive $Q^2$'s increases fast when the order index $M$ 
of the Pad\'e approximant increases. When $Q^2$'s are complex,
the precision range of $|Q^2|$'s decreases when $Q^2$ approaches
the singularity cut.
The analytic analogs ${\mathcal{A}}_n^{(mMA)}$ of higher powers 
$a^n = (\alpha_s/\pi)^n$ ($n \geq 2$) are then evaluated
as combinations of logarithmic derivatives of the approximant
$R_{M}^{M-1}(Q^2)$ [$\approx {\mathcal{A}}_1(Q^2)$].
The approximants obtained in this way for ${\mathcal{A}}_n^{(mMA)}$
show less precision when $n$ increases
and/or when $Q^2$ approaches the singularity cut.
These approximants still work fine if we increase
the order index $M$ of the Pad\'e. However, high precision
is needed only for the $n=1$ case, because
the higher couplings ${\mathcal{A}}_n^{(mMA)}$
get strongly suppressed (even at low $|Q^2|$) in analytic
QCD when $n$ increases. Further, when evaluation of observable
involves contour integration (such as, for example, in the case of $r_{\tau}$),
the contributions of ${\mathcal{A}}_n(Q^2)$ get supressed by
the rest of the integrand when $Q^2$ comes close to the singularity cut.
While we generally used for the IR cutoff $\sigma_{\rm cut}$ ($\sim M_{\pi}^2$)
the specific value $\sigma_{\rm cut} (\equiv M_0^2) = 4 M_{\pi}^2$,
we also showed that the conclusions in this work
are independent of the specific value chosen, 
by comparing various results for $M_0 = 2 M_{\pi}$ and
$M_0 = 3 M_{\pi}$.

We further suggested an inclusion of one additional Dirac
delta function to the mMA discontinuity function 
$\theta(\sigma-M_0^2) \rho_1(\sigma)$
at low energies where the precise behavior of $\rho_1$
is unknown - Dirac modified MA model (dmMA). 
Such a modification maintains the analyticity
at $Q^2=0$ and its vicinity, and 
allows us to reproduce the experimental value of the
semihadronic $\tau$ decay ratio $r_{\tau}$.
Such a modification keeps the same form of the
Pad\'e approximants $R_{M}^{M-1}(Q^2)$ of ${\mathcal{A}}_1^{(dmMA)}$
as in the mMA case and allows us to evaluate them (and the higher power analogs)
in an easy and efficient manner.
The (three) parameters of such a model can be determined
by requiring that the model reproduces the measured values
of low energy QCD observables \cite{wp}. 

\begin{acknowledgments}
This work was supported by FONDECYT Grant No. 1095196 (G.C.) and a
PIIC-USM grant (H.M.)
\end{acknowledgments}

\end{document}